\documentclass[onecolumn,groupedaddress,showkeys]{article}
\usepackage{float}
\usepackage{graphicx}
\usepackage{amssymb}
\usepackage{authblk}
\usepackage{amsmath}
\usepackage{caption}
\usepackage{subcaption}
\usepackage{xcolor}
\usepackage[super]{cite} 
\usepackage[breaklinks]{hyperref}
\hypersetup{colorlinks,urlcolor=black,citecolor=black,linkcolor=black,filecolor=black}
\usepackage{breakurl}
\usepackage{anysize}
\usepackage[left=1cm,right=1cm,top=1cm,bottom=1cm,includehead,includefoot]{geometry}
\begin{document}

	\title{Cellular automata in the light of COVID-19}
	\author{Sourav Chowdhury\thanks{email: chowdhury95sourav@gmail.com}\qquad Suparna Roychowdhury\thanks{email: suparna@sxccal.edu}\qquad Indranath Chaudhuri\thanks{email: indranath@sxccal.edu}}
	\affil{Department of Physics, St. Xavier's College (Autonomous)\\
		30 Mother Teresa Sarani, Kolkata-700016, West Bengal, India}

	\maketitle
	
\begin{abstract}		
	 Currently, the world has been facing the brunt of a pandemic due to a diseases called COVID-19 for the last two years. To study the spread of such infectious diseases it is important to not only understand their temporal evolution but also the spatial evolution. In this work, the spread of this disease has been studied with a cellular automata (CA) model to find the temporal and the spatial behavior of it. Here, we have proposed a neighborhood criteria which will help us to measure the social confinement at the time of the disease spread. The two main parameters of our model are (i) disease transmission probability ($q$) which helps us to measure the infectivity of a disease and (ii) exponent ($n$) which helps us to measure the degree of the social confinement. Here, we have studied various spatial growths of the disease by simulating this CA model. Finally we have tried to fit our model with the COVID-19 data of India for various waves and have attempted to match our model predictions with regards to each wave to see how the different parameters vary with respect to infectivity and restrictions in social interaction.
\end{abstract}
	
\section{Introduction}
	Epidemics and pandemics have a long story throughout human history. Recently human civilization has faced another pandemic named COVID-19. This pandemic has affected many countries through multiple waves. Total of 504,451,689 people have been infected worldwide and 6,222,430 people have died due to COVID-19 till 17 April 2022. In India 43,042,097 people have suffered and 521,781 have died as of 17/04/2022 due to this disease \cite{worldometers}. COVID-19 is caused by the virus which is named SARS COV-2. Multiple variants of this virus, like delta, omicron and many others makes it harder to control and predict its behavior. Recently another variant of COVID-19 named XE has been found \cite{WHO}. Mathematical modeling helps us to understand the behavior of disease spread such that prevention and control strategies can be built. Also, mathematical models can help us to find some inherent properties of the disease and nature of its spread.
	
	There are many different types of models that have been used in the past to study various diseases. These models are mainly modified versions of the Kermack McKendrick SIR model which is based on a system of coupled ordinary differential equations \cite{basic_SIR}. Currently, the ODE-based models and statistical models are widely used in literature to model the temporal behavior of the spread of COVID-19 from different aspects. Most of these models have tried to analyze the spread of this disease and tried to predict its future behavior  \cite{third_wave_model_stat,third_wave_model,kavitha2021second,gowrisankar2022omicron}. There are models which have proposed various intervention and vaccination strategies to prevent and control the spread of the disease \cite{covid_vac_india,covid_vac_acc_india,NPI_vs_vac,vac_resistant,3rdwave_covid}. Also, some authors have tried to predict different inherent properties of this pandemic like herd immunity and its chaotic nature \cite{HIT, gowrisankar2020can,chaos_vac,chaos_vac_osc, easwaramoorthy2021exploration}. These temporal models can give us much valuable information, however most of these models assume that a population is homogeneously mixed and cannot describe any spatial behavior. To incorporate this spatial behavior, deterministic and probabilistic Spatio-temporal models have been used in recent studies. Cellular automata (CA) is one such kind of spatio-temporal model.
	
	Cellular automata (CA) has been used in many studies to model different aspects of epidemics. It has been widely used to model the disease spread of influenza and various vector-borne diseases like dengue \cite{Bubonic_plague_CA, influenza_seir_reg1, influenza_egypt, CA_influenza_abu_dhabi, influenzaA_SLEIRD, chikungunya_vector_reg2, dengue_vector, dengue_CA,  Ebola_CA}. A neighborhood condition is an important aspect in the CA. The most used neighborhood conditions are (i) Neumann's neighborhood condition, (ii) Moore's neighborhood condition, (iii) Extended neighborhood condition, and (iv) Random interactions. Coupled with these neighborhood conditions, various models like SEIR, SEIRS, SEIRD, and SEIRQD have been studied with the help of CA to model the spatial growth of epidemics \cite{influenza_seir_reg1, influenzaA_SLEIRD,  CA_nghbd_sat, CA_small_world, CA_osc_beta, CA_SEIRS, CA_spt_heter, two_reg_CA}. Currently, CA has gained a lot of momentum in the studies of COVID-19. Various advanced studies with Genetic algorithms and network models have been for COVID-19 data. \cite{covid_social_isol, covid_vac_lockdwn, covid_GA_1, covid_GA_2,covid_mob_restr_net, k_index_net}. Also there are models where COVID-19 has studied from different aspects.  
	
	In this paper, we have tried to model COVID-19 using cellular automata (CA) to find spatio-temporal behavior of it. We have also made some analysis to understand its behavior in different waves of the disease.  A cellular automata (CA) model is represented in a square lattice and defined by some neighborhood and boundary conditions, the details of which are given in following sections.
	
	This paper is arranged as follows: Section~\ref{math_mod_sec} consists of a detailed discussion of the model, neighborhood conditions, probability of infection, and the algorithm of the model. The result of the simulation has been shown in section~\ref{sim_sec} and the data analysis is shown in section~\ref{data_analysis}. Finally, Section~\ref{conclusion} consists of the conclusions of our model. 
\section{Mathematical Model}\label{math_mod_sec}
	In this article, we have illustrated a Cellular automata (CA) model for epidemics and assumed the SEIR model as the base model. SEIR model stands for Susceptible-Exposed-Infectious-Removed. Here we have considered a $N\times N$ square lattice, where each cell of the lattice is assumed to be a person. Each cell of the lattice can have the set of states, $\mathcal{S}=\left\{S, E, I, R\right\}$ and these states are represented by the values $\mathcal{V}=\left\{0, 1, 2, 3\right\}$. The updation of a cell's state depends upon various conditions like, (i) the current state of the cell, (ii) the amount of time spent in the current state, and (iii) the current states of the neighbors. The main assumptions of our model are given below:
		
	\begin{itemize}
		\item Every cell represents a person.
		\item Only susceptible persons can interact with the other cells.
		\item A removed person cannot be infected again.
		\item For this CA model, we have assumed a periodic boundary condition. If a cell of $i$th row and $j$th column of a $N\times N$ lattice is denoted by $(i,j)$ then,
		\begin{equation}
			\begin{split}
				(N+1,j)\equiv (1,j)\qquad j=0,1,...N.\\
				(i,N+1)\equiv (i,1)\qquad i=0,1,...N.\\
			\end{split}
		\end{equation}
		\item One susceptible person can interact with a single person in each time step.
	\end{itemize}
	
	\subsection{Neighborhood condition}\label{nghbd_cond}
		Nearest neighborhood condition is a widely used concept in the literature. Here, it has been assumed that a particular cell can only interact with its nearest neighborhood cells. Such two famous neighborhood conditions are: (i) Neumann's neighborhood condition and (ii) Moore's neighborhood condition.
		
		\begin{figure}[H]
			\begin{subfigure}{.49\textwidth}
				\centering
				\includegraphics[scale=0.5]{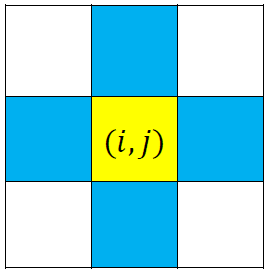}
				\caption{Neumann's neighborhood condition.}
				\label{nghbd_Neu}
			\end{subfigure}
			\begin{subfigure}{.49\textwidth}
				\centering
				\includegraphics[scale=0.5]{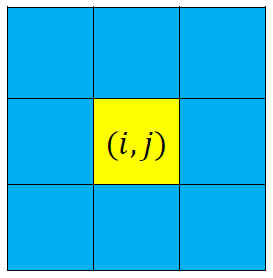}
				\caption{Moore's neighborhood condition.}
				\label{nghbd_M}
			\end{subfigure}
			\caption{{Different neighborhood conditions. Blue cells represent the nearest neighborhoods of $(i,j)$ cell.}\label{nghbd_gen}}	
		\end{figure}
		
		Fig.~\ref{nghbd_gen} shows the two neighborhood conditions. Fig.~\ref{nghbd_Neu} shows Neumann's neighborhood condition, where the nearest neighborhoods of any chosen $(i,j)$ cell are the first neighborhood cells with respect to the chosen cell. Similarly, Fig.~\ref{nghbd_M} shows Moore's neighborhood condition. In this case, all first and second neighborhoods are treated as the nearest neighborhoods of the chosen $(i,j)$ cell.
		
\begin{figure}[H]
				\centering
				\includegraphics[scale=0.55]{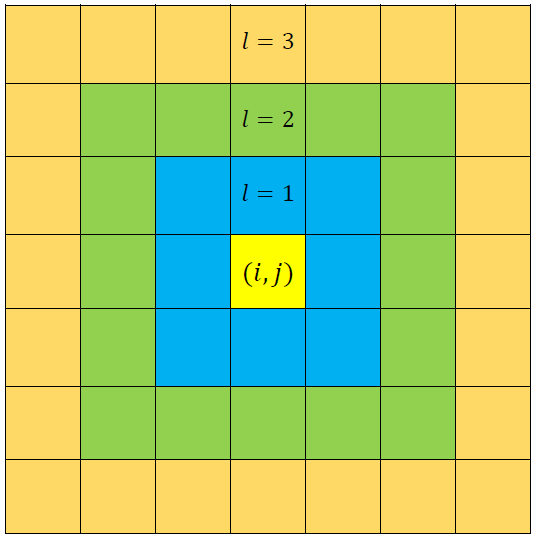}
				\caption{{Different layers of a lattice with respect to $(i,j)$ cell.}\label{nghbd_custom}}
			\end{figure}

		In this work, we have assumed that a cell can interact with any other cells depending on the probability of interaction ($p_{int}$) between them. Here we have assumed that the probability of interaction  ($p_{int}$) of a cell $(i,j)$ to any other cell varies inversely as a function of $d$ (distance between two cells) in the form of a power law. Hence,
		
		\begin{equation}
			p_{int}(d)\propto \frac{1}{d^{n}} \label{pint_d}
		\end{equation}
		where $n$ is the degree exponent and can have a value greater than zero. Here, we have assumed that the distance between two cells is not just the geometrical distance between two. It depends on the layer number ($l$). In Fig.~\ref{nghbd_custom}, we have shown how layers are defined. Also, it shows that a layer $l$ contains the $8l$ number of cells. If we choose a lattice of size $N\times N$ then the total number of layers in this lattice is $L=\frac{N-1}{2}$, when $N$ is odd.
			
			Hence, we can write,
			
			\begin{equation}
				d\propto l.
			\end{equation}
			
			For mathematical simplicity, we can assume $d=l$. Hence, from Eq.~\ref{pint_d} we can write,
			
			\begin{equation}
					p_{int}(d)=p_{int}(l)\propto \frac{1}{l^{n}}\label{pint_l}
			\end{equation}
		
			If there are $L$ number of layers then the above equation can be written as,
			
			\begin{equation}
				p_{int}(l)=\frac{\frac{1}{l^{n}}}{\sum_{l=1}^{L}\frac{1}{l^{n}}}=\frac{1}{A_{n}l^{n}}\label{pint_main}
			\end{equation}
		
			where, $A_{n}=\sum_{l=1}^{L}\frac{1}{l^{n}}$. Hence, a person at the $(i,j)$ cell can interact with any other cell of layer $l$ with a probability $p_{int}(l)$. Thus, average interaction distance ($\langle d\rangle$) can be defined as, 
			\begin{equation}
				\langle d\rangle=\sum_{l=1}^{L}lp_{int}(l)=\frac{1}{A_{n}}\sum_{l=1}^{L} l \frac{1}{l^{n}}=\frac{1}{A_{n}}\sum_{l=1}^{L} \frac{1}{l^{n-1}}
			\end{equation}
		
			\begin{figure}[H]
				\begin{subfigure}{.5\textwidth}
					\centering
					\includegraphics[scale=0.35]{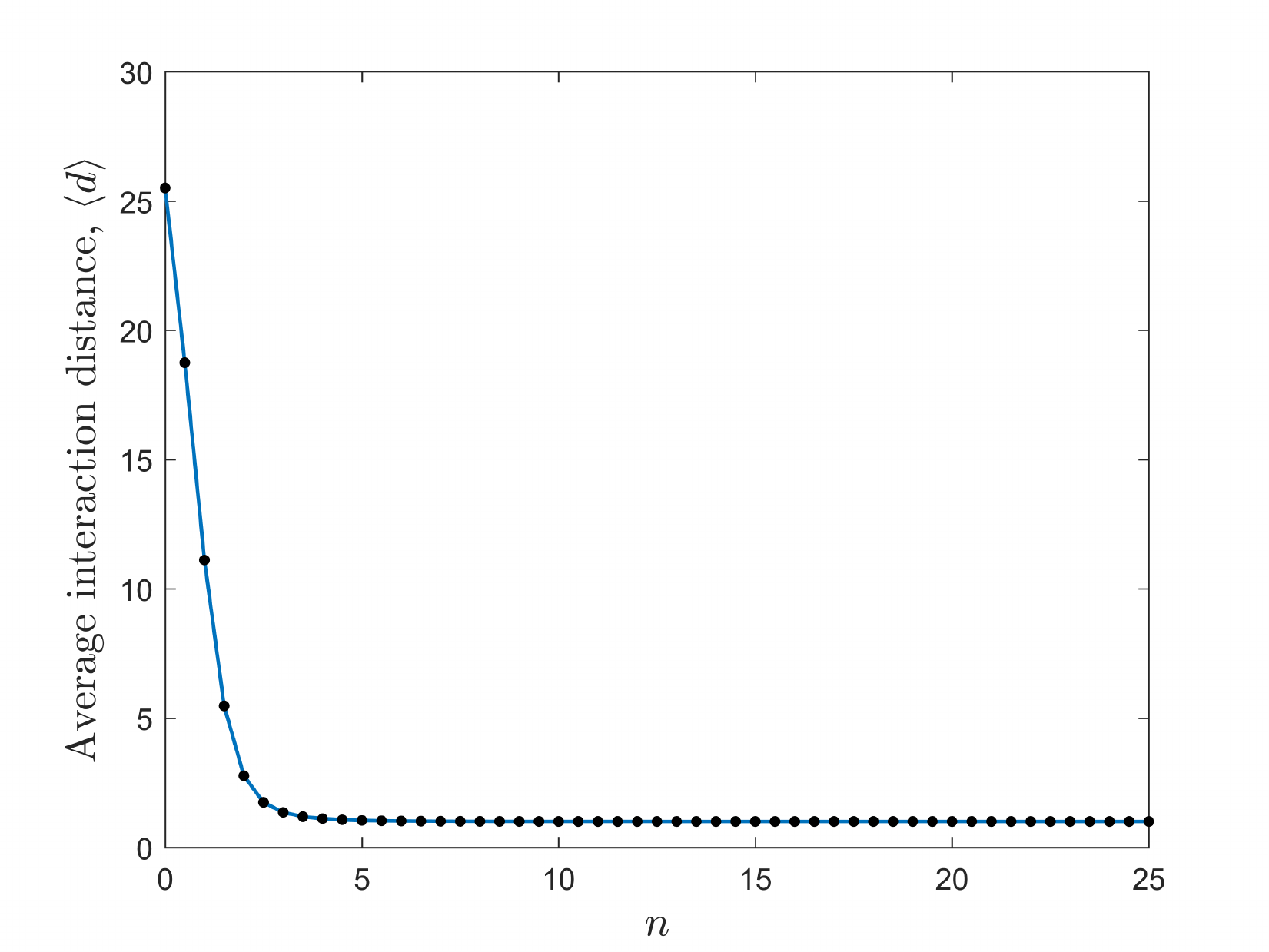}
					\caption{}
					\label{avg_int_n}
				\end{subfigure}
				\begin{subfigure}{.5\textwidth}
					\centering
					\includegraphics[scale=0.35]{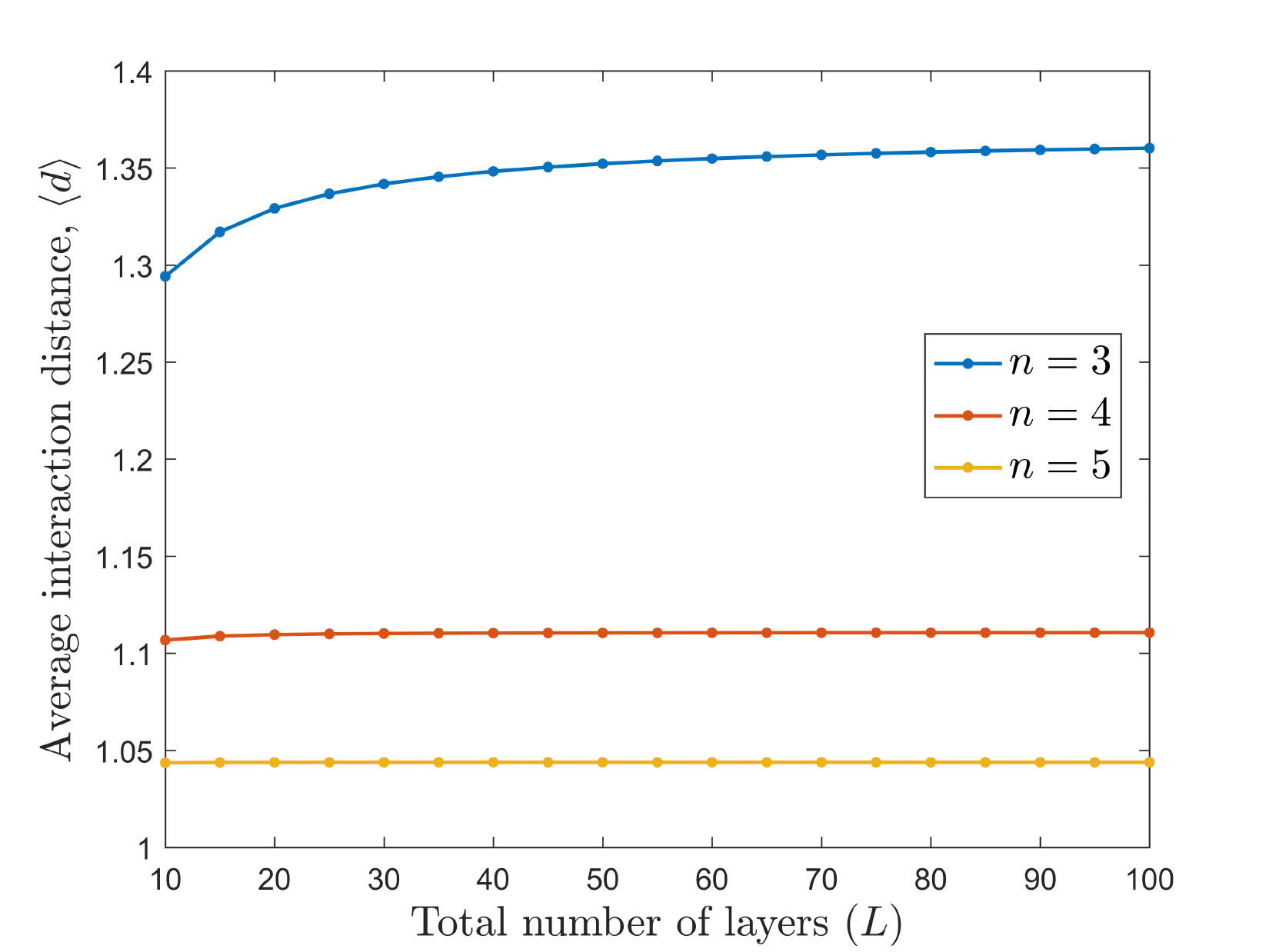}
					\caption{}
					\label{avg_int_L}
				\end{subfigure}
				\caption{{Plots of average interaction distance ($\langle d\rangle$) with $n$ and the total number of layers ($L$). (a) Plot of average interaction distance ($\langle d\rangle$) with $n$ by considering the total number of layers ($L$) = 50. (b) Plot of average interaction distance ($\langle d\rangle$) with the total number of layers ($L$) for $n$=3,4, and 5. }\label{avg_int}}	
			\end{figure}
			
			Fig.~\ref{avg_int_n} shows the variation of average interaction distance ($\langle d\rangle$) with the degree exponent ($n$). From this figure, it can be found that the average interaction distance ($\langle d\rangle$) is quickly decreases and saturates to unity as $n$ increases. Also, from Fig.~\ref{avg_int_L} it can be seen that the average interaction distance ($\langle d\rangle$) is approximately $\sim 1$ for exponents $n>3$. Hence for $n\gg 3$ the neighborhood condition is approximately similar to Moore's neighborhood condition as discussed earlier and does not give any significantly different results.
	\subsection{Probability of infection ($Q_{I}$)}
		Let, $q$ denote the disease transmission probability when a susceptible and an infectious person interact. The probability that a susceptible person will interact with any person at the layer $l$ is $p_{int}(l)$. If the probability of finding an infectious person in that layer is $p_{I}(l)$ then the probability that the susceptible person will be infected is $qp_{int}(l)p_{I}(l)$. Hence, the probability of infection ($Q_{I}$) of a susceptible person is,
		
		\begin{equation}
			Q_{I}= q \sum_{l=1}^{L} p_{int}(l)p_{I}(l). \label{prob_inf1}
		\end{equation}	
		As, $p_{int}(l)=\frac{1}{A_{n}l^{n}}$, from the above equation (Eq.~{\ref{prob_inf1}})	we can write,
		\begin{equation}
			Q_{I}= \frac{q}{A_{n}} \sum_{l=1}^{L} \frac{p_{I}(l)}{l^{n}} \label{prob_inf2}
		\end{equation}		
		
		From the above equation (Eq.~\ref{prob_inf2}) we can say that the terms with small layer number ($l$) dominate the summation. Hence, the infection possibility of a susceptible person mainly depends on the infection situation around the person. 
		
		Thus in our model, instead of choosing a traditional neighborhood condition where the degree of the interaction is fixed, we have assumed a model where we can vary the degree of the social confinement by changing $n$ (degree exponent). Also we have calculated the probability of infection ($Q_{I}$) for this modified model.
\section{Algorithm and Simulations} \label{sim_sec}
	\subsection{Algorithm}
	Here, we have discussed the state updation algorithm of the SEIR model. As we have mentioned earlier, every cell's state is denoted by a value ($v$) which is present in this set $\left\{0, 1, 2, 3\right\}$. The algorithm is given below:
	
	\begin{itemize}
		\item Let, at time $t$ there is a susceptible person at $(i,j)$ cell. So, the value of the $(i,j)$ cell is $v(i,j,t)=0$ and the probability of infection is $Q_{I}(i,j,t)$. To find the infection possibility of the susceptible person, we will generate a uniform random number $u$ between 0 and 1. 
		
		If, $u\leq Q_{I}(i,j,t)$ then the susceptible person is exposed and at time $t+1$ the state of the $(i,j)$ cell will be changed from $v=0$ to $v=1$.
		
		else, at time $t+1$ the state of the $(i,j)$ cell will be unchanged.
		
		\item An exposed person ($v=1$) will remain exposed for $\tau_{I}$ number of days. After that, the person will be infectious and the state of the corresponding cell will be changed from $v=1$ to $v=2$.
		
		\item An infectious person will remain in this state for $\tau_{R}$ number of days. After that, the person will be removed (recovered or dead) and the state of the cell will be changed from $v=2$ to $v=3$.
	\end{itemize}
	\subsection{Simulation}
	In this part, we have done simulation of our model with $n=1,2,3$. The values of the parameters and initial conditions that are used in the simulations are listed in the tables (Table~\ref{par_tab} and Table~\ref{init_tab}) below:
	\begin{table}[H]
		\centering
		\begin{tabular}{|l|c|c|}
			\hline
			Description of the parameters & Parameters & Values of the parameters\\      
			\hline
			Lattice size & $N\times N$ & $101\times 101$\\
			Disease transmission probability & $q$ & 0.3\\
			Latency period of the disease & $\tau_{I}$ & $8~days$ \\   
			Removal period & $\tau_{R}$ & $18~days$ \\ 
			\hline                     
		\end{tabular}
		\caption{{Table for the parameter values that are used in the simulations.}\label{par_tab}}
	\end{table}
	\begin{table}[H]
		\centering
		\begin{tabular}{|c|c|}
			\hline
			States & Initial values\\
			\hline
			$S(0)$ & 10200 \\
			$I(0)$ & 1 \\
			$E(0)$ & 0\\
			$R(0)$ & 0 \\
			\hline                     
		\end{tabular}
		\caption{{Table for the initial conditions of the simulations.} \label{init_tab}}
	\end{table}
	\begin{figure}[H]
		\begin{tabular}{ccc}
			\centering
			\includegraphics[scale=0.22]{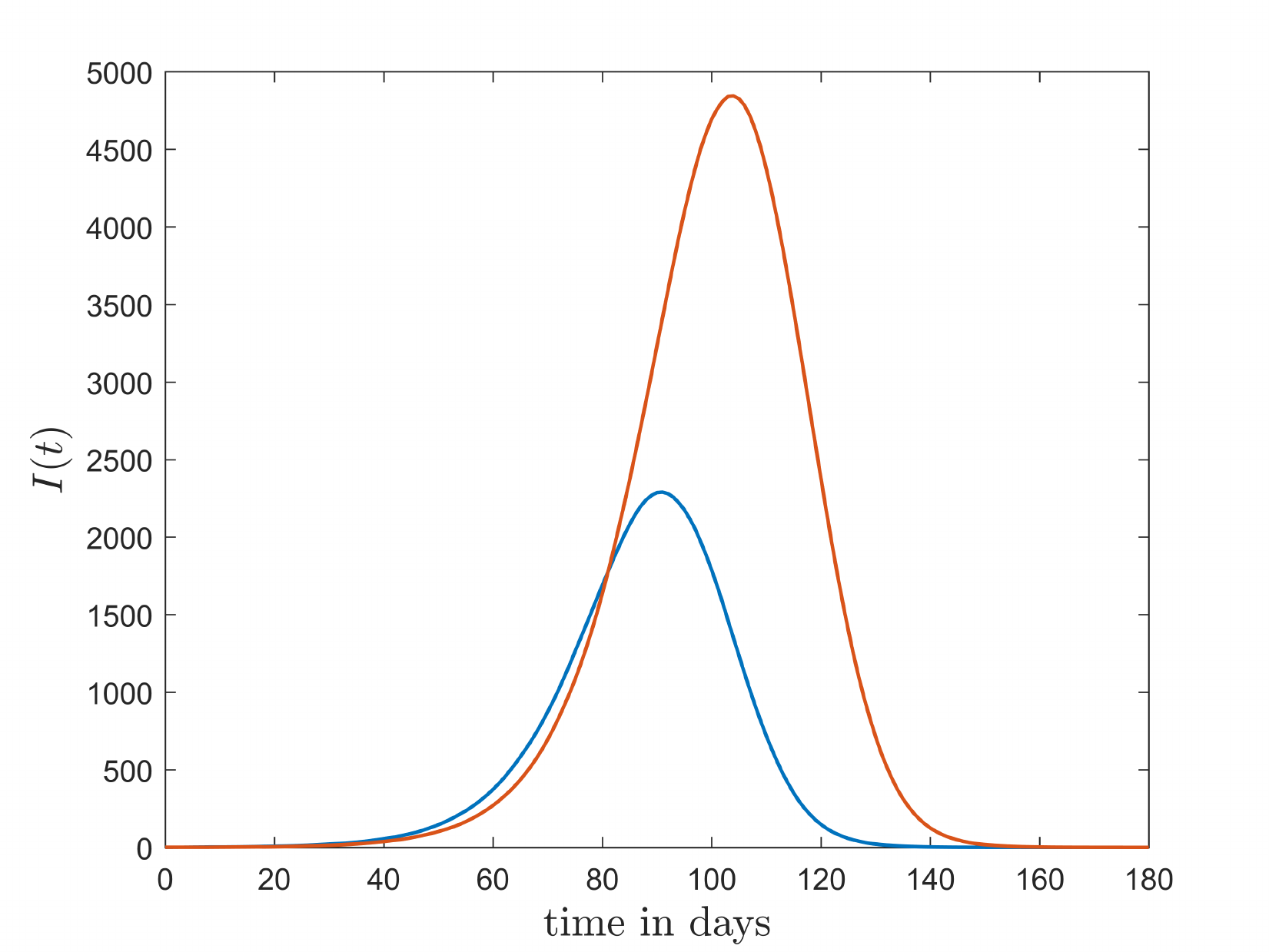}&
			\includegraphics[scale=0.22]{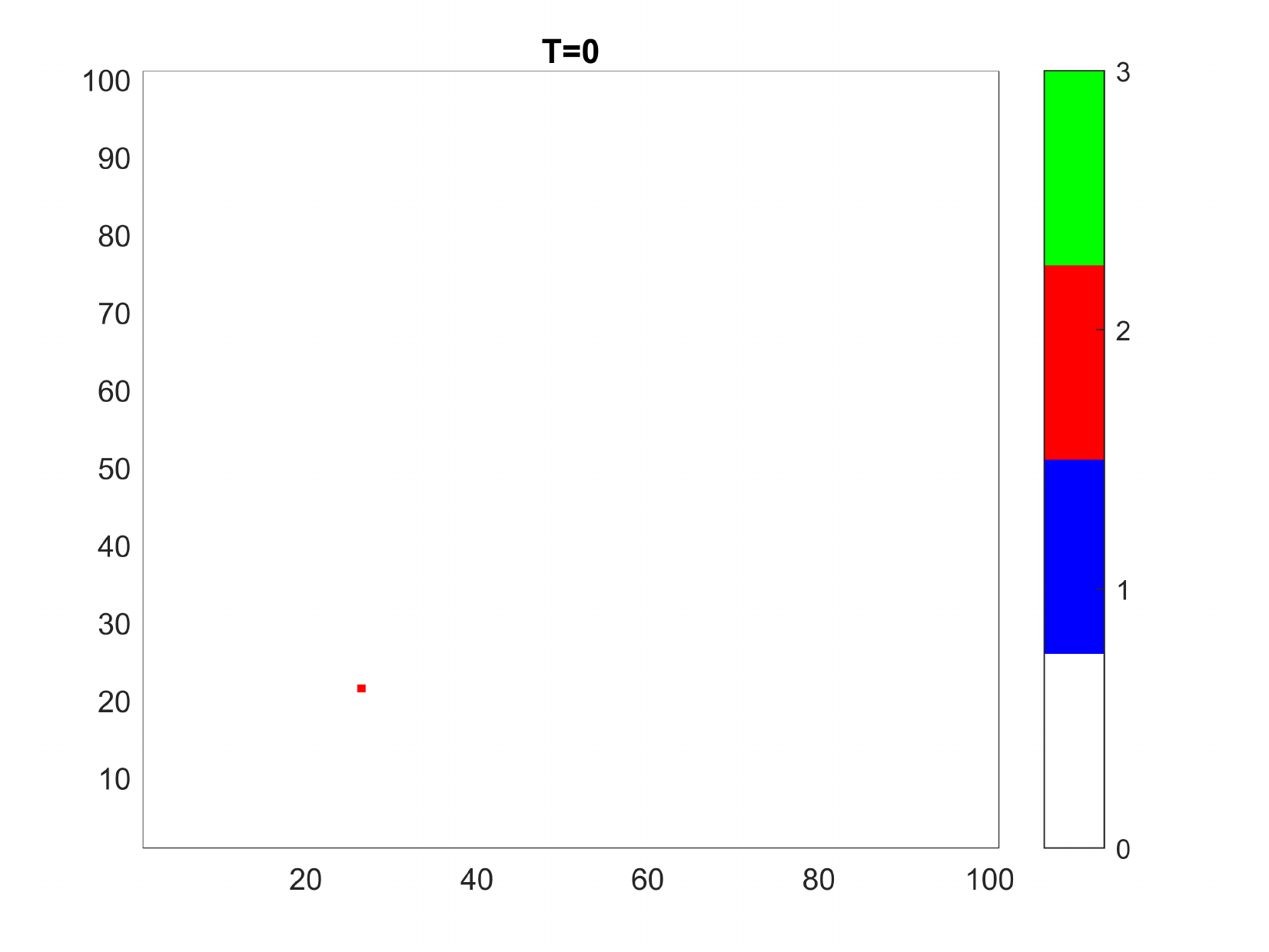}&
			\includegraphics[scale=0.22]{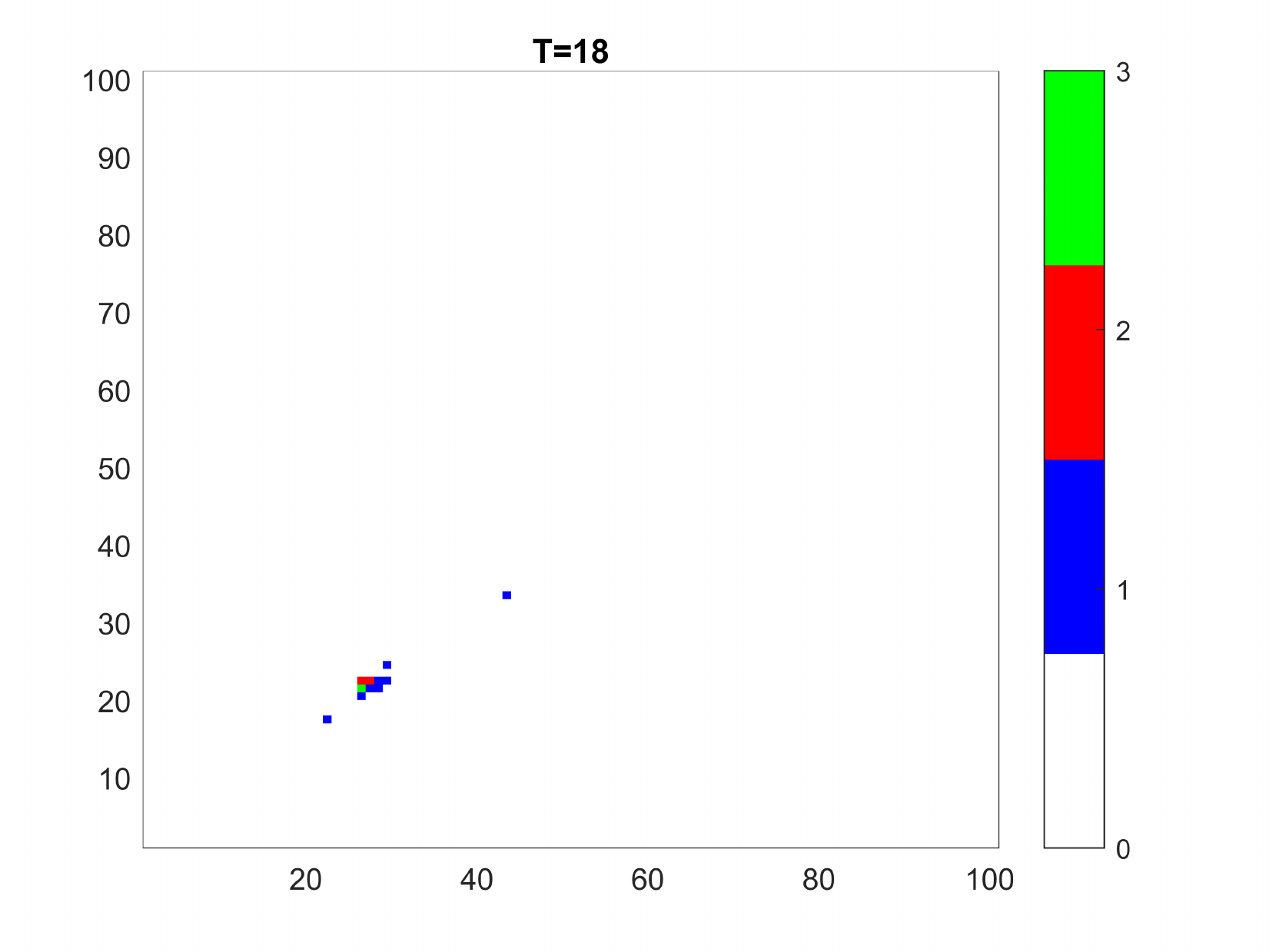}\\
			\includegraphics[scale=0.22]{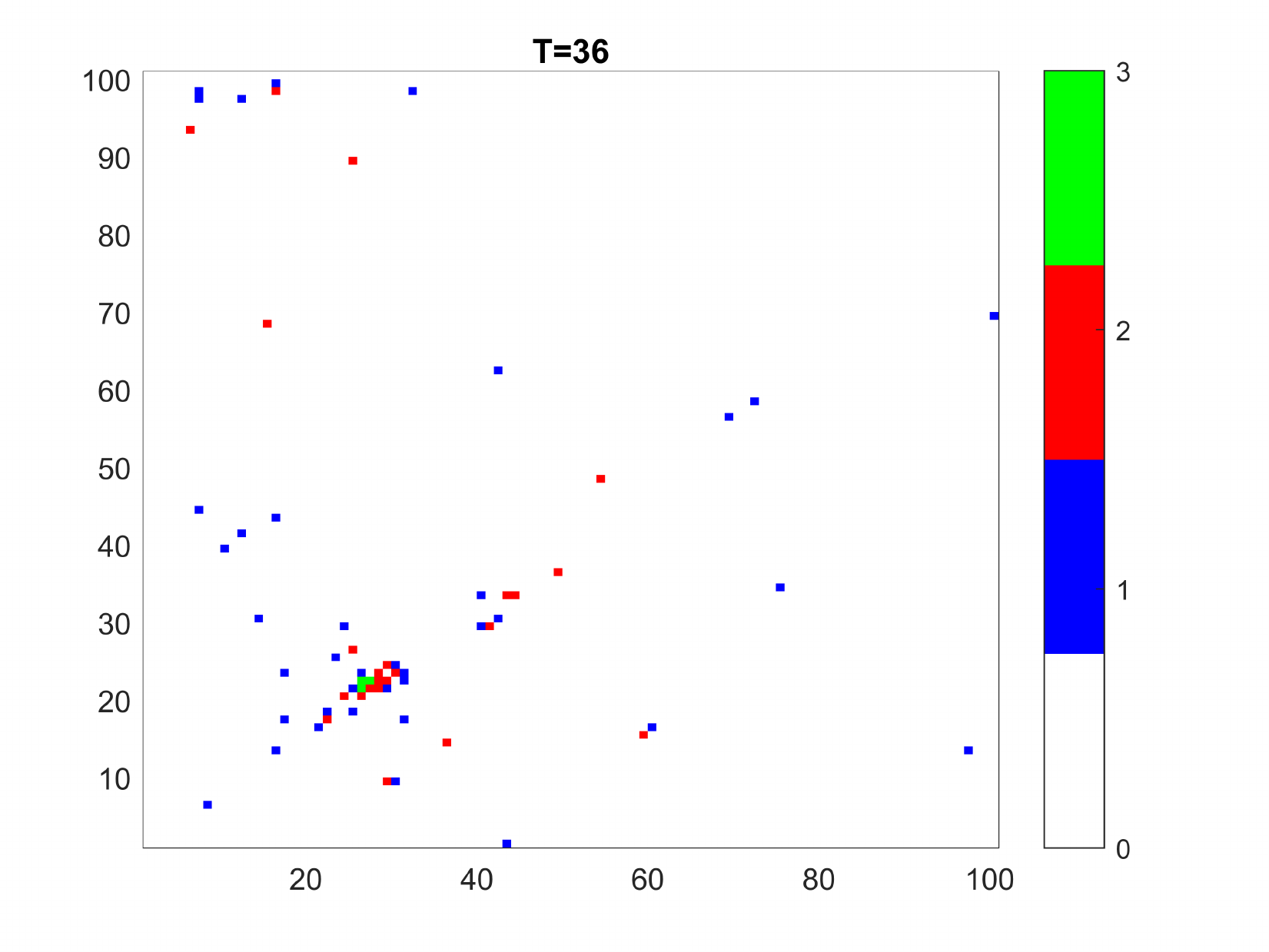}&
			\includegraphics[scale=0.22]{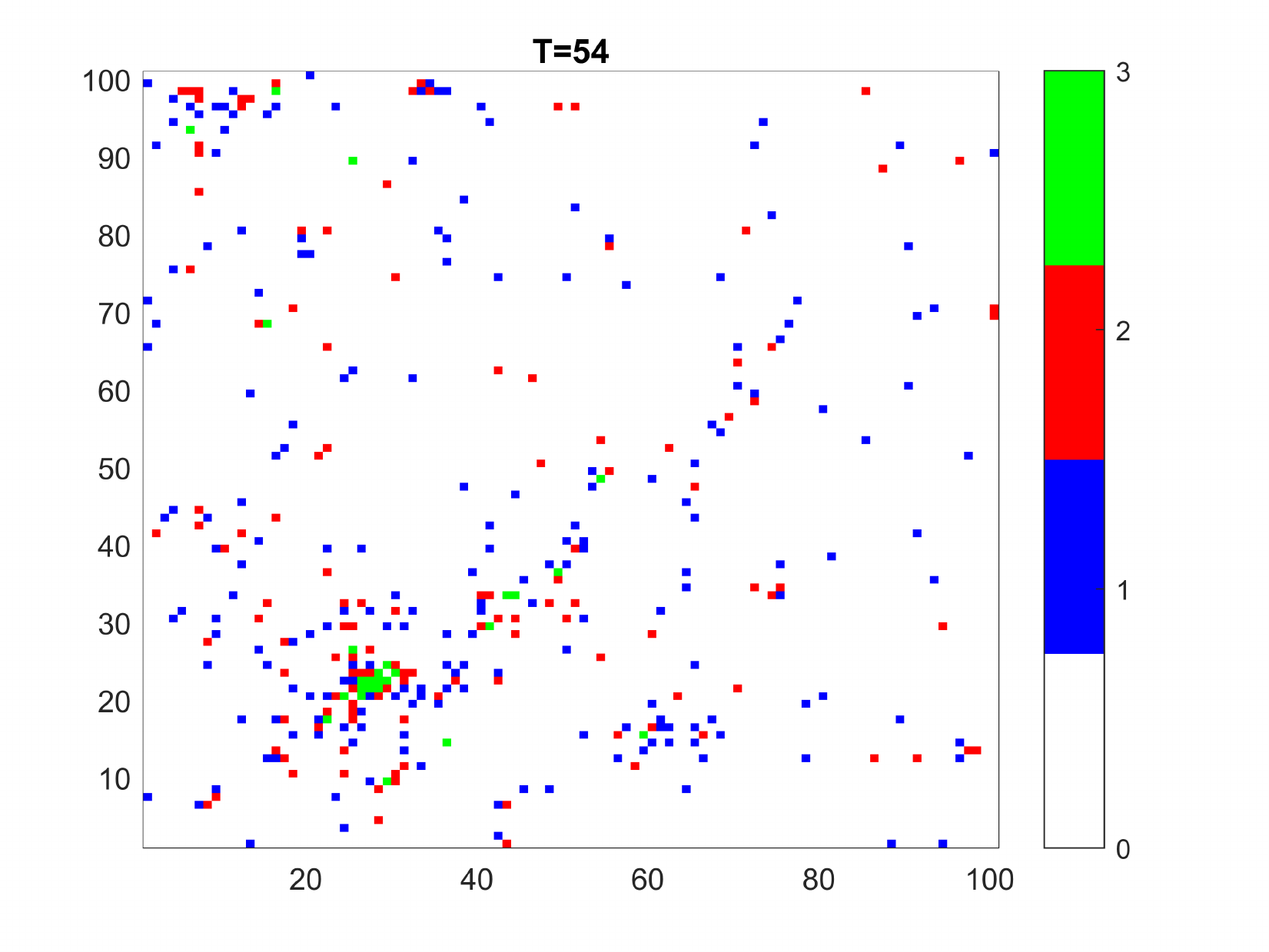}&
			\includegraphics[scale=0.22]{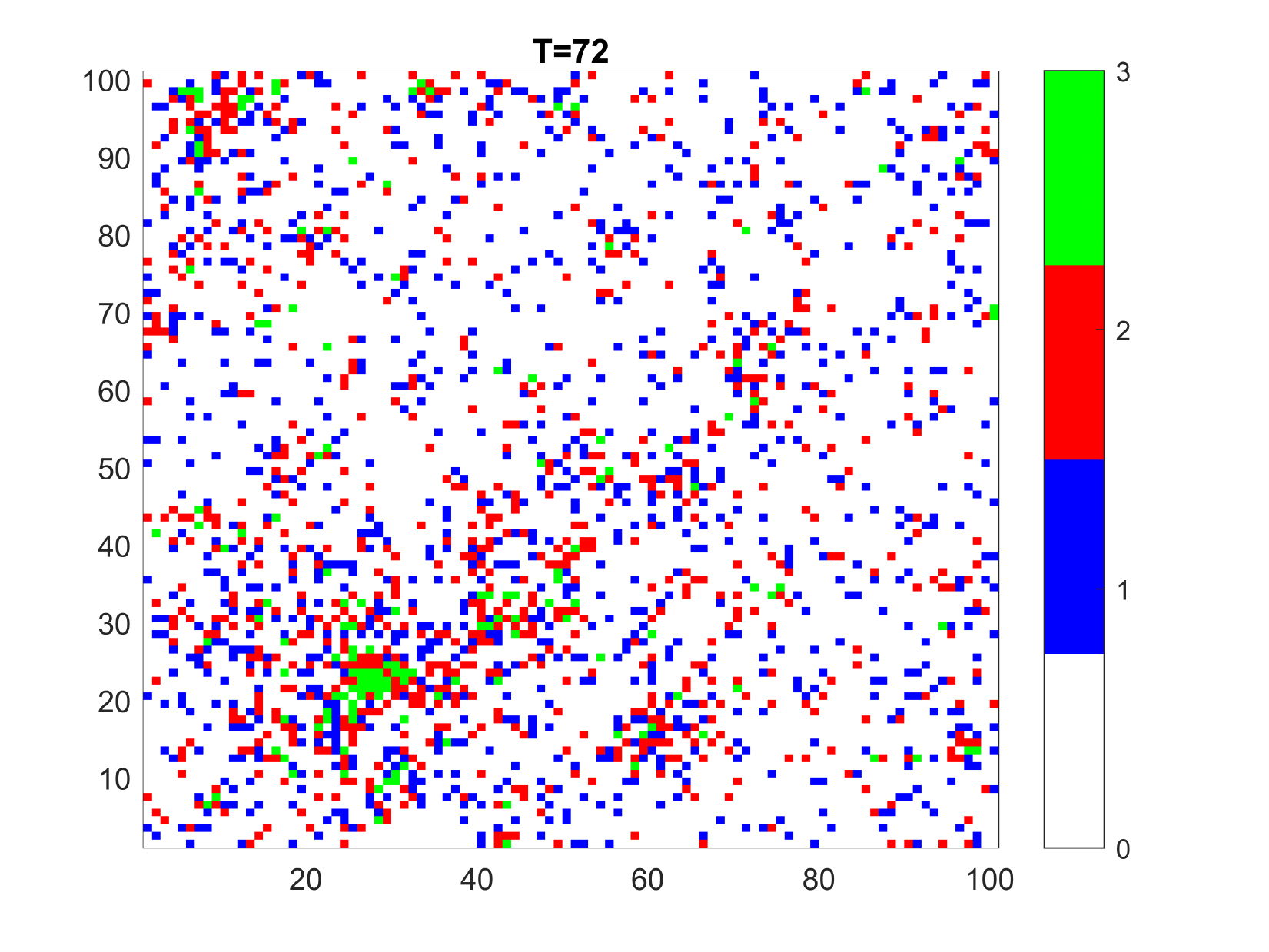}\\
			\includegraphics[scale=0.22]{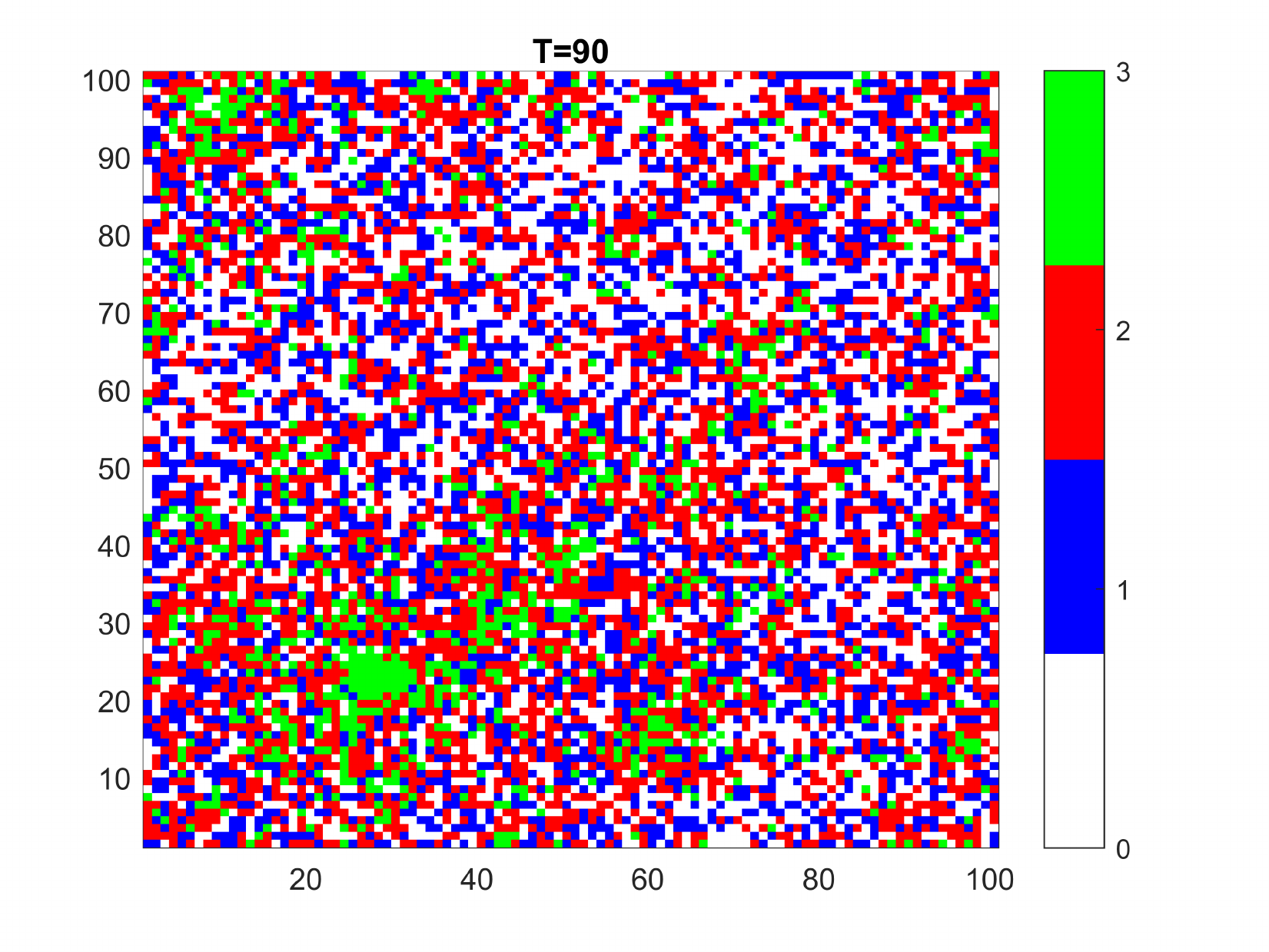}&
			\includegraphics[scale=0.22]{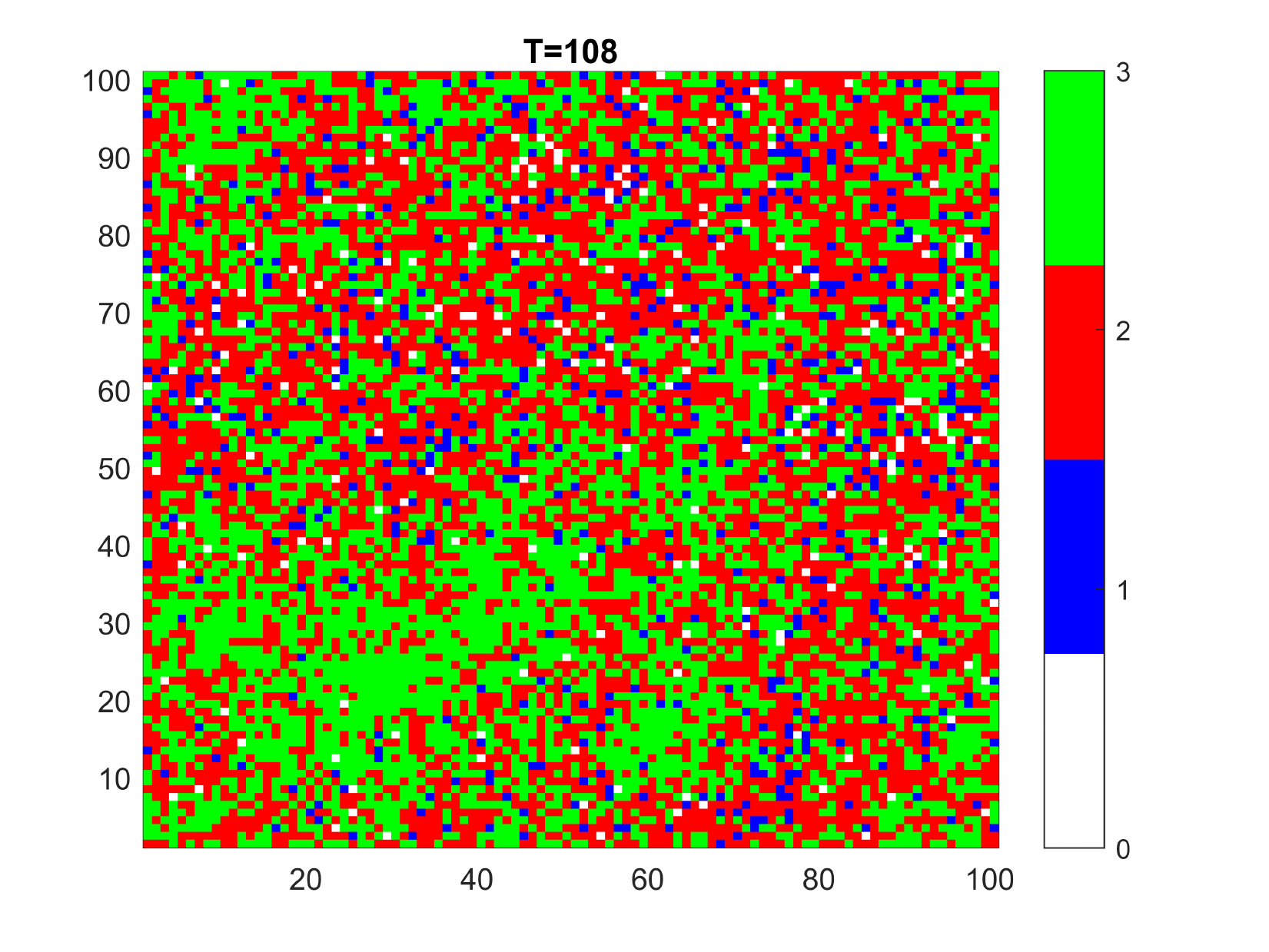}&
			\includegraphics[scale=0.22]{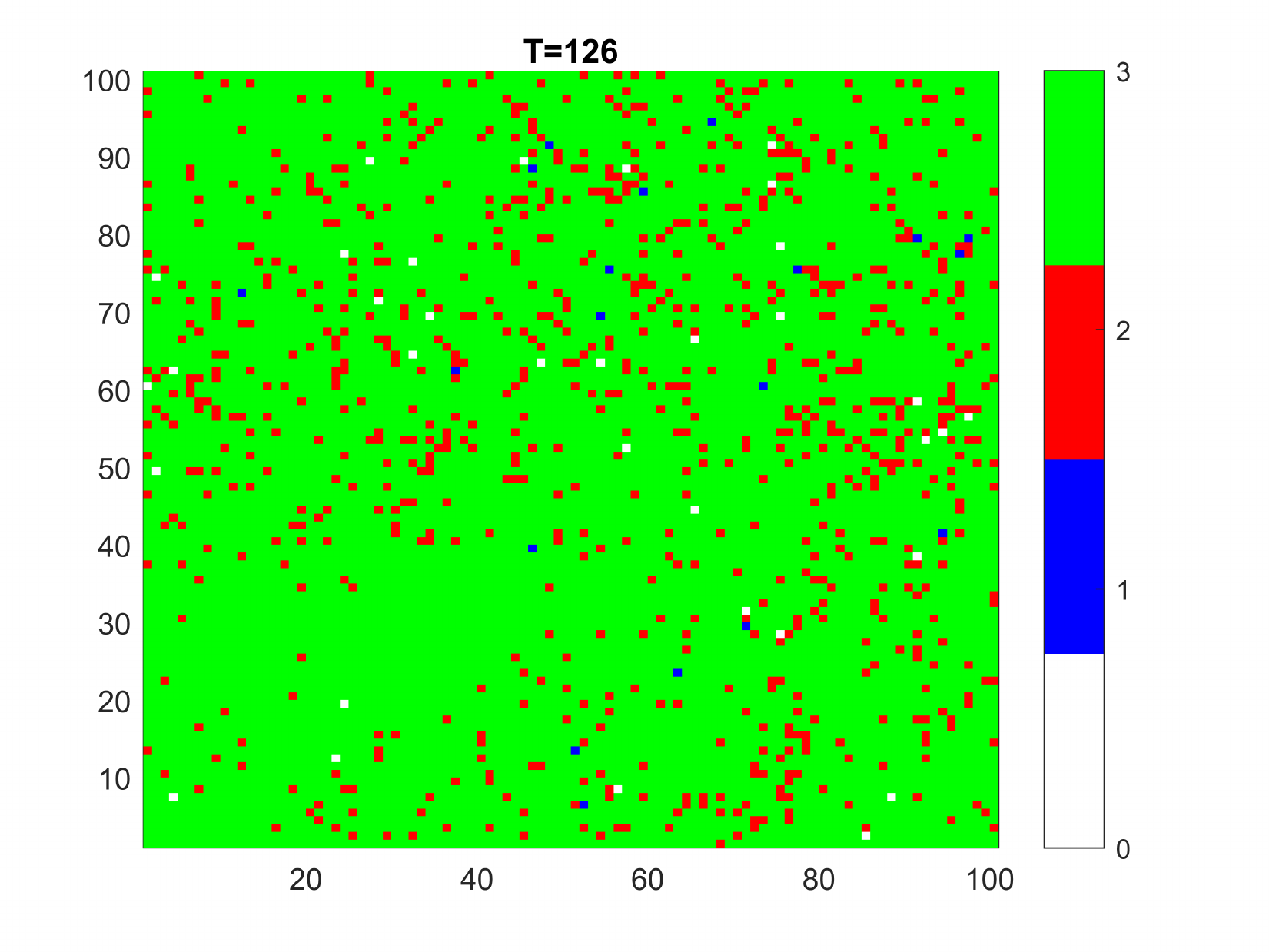}\\
			\includegraphics[scale=0.22]{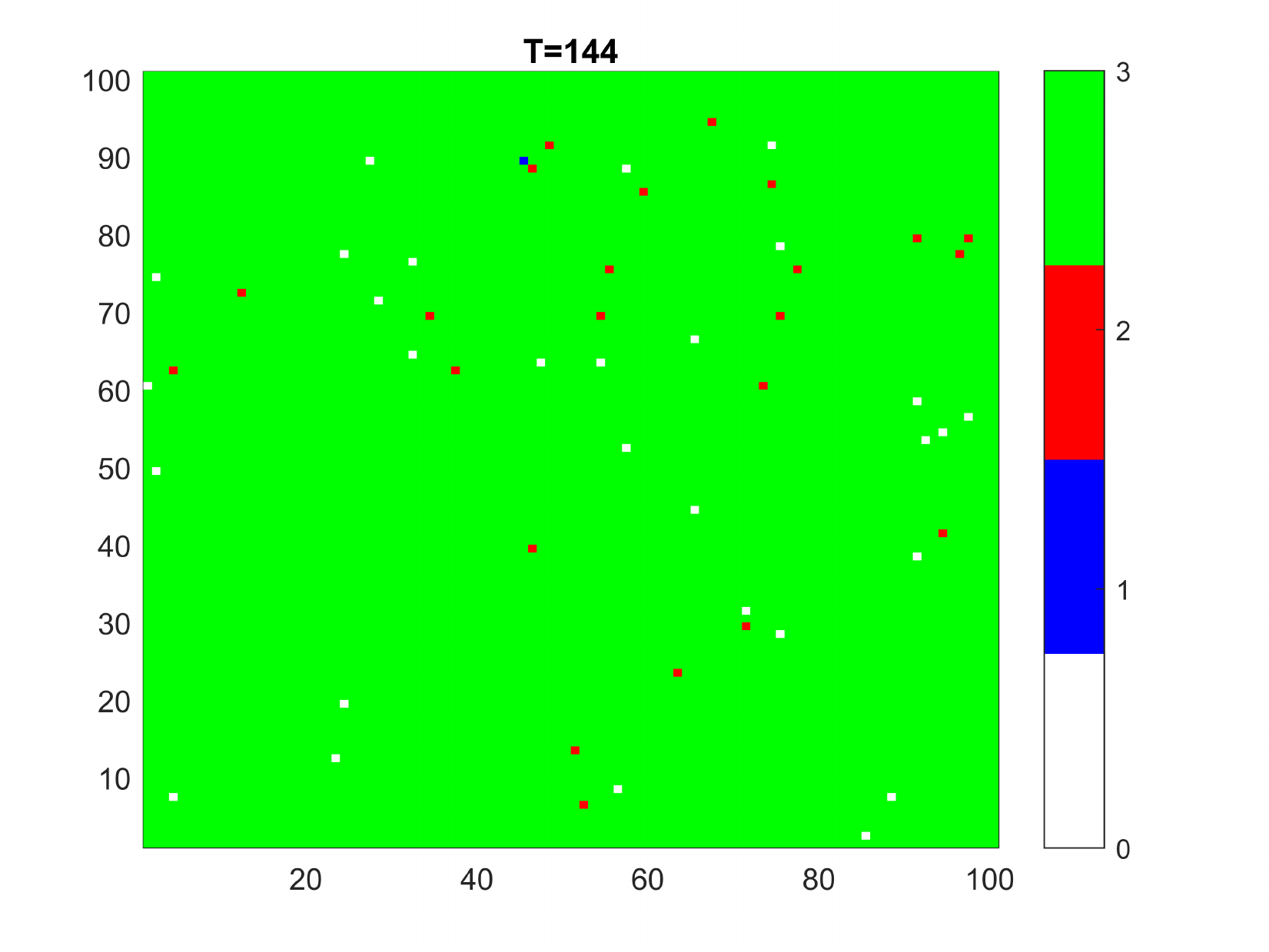}&
			\includegraphics[scale=0.22]{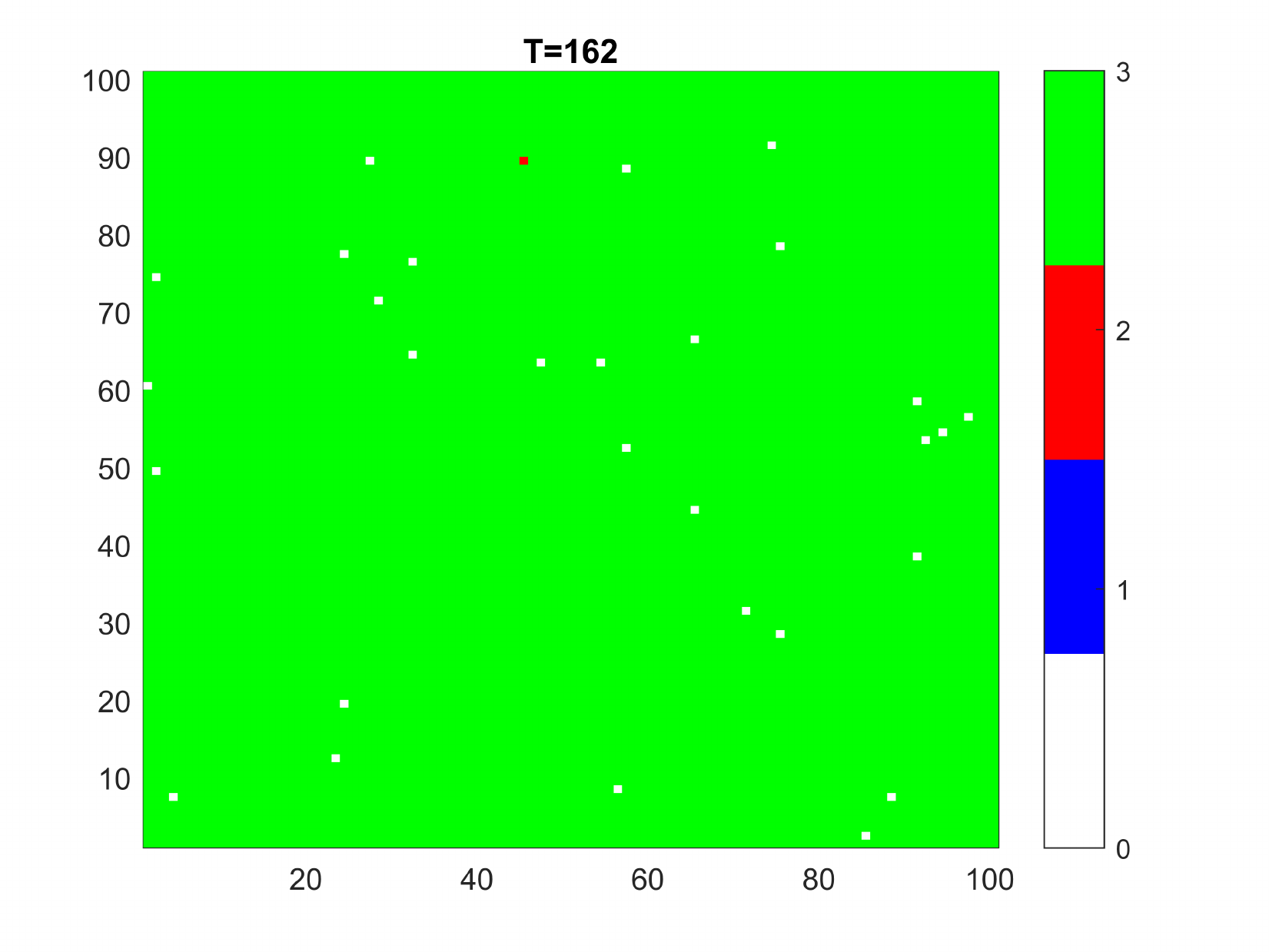}&
			\includegraphics[scale=0.22]{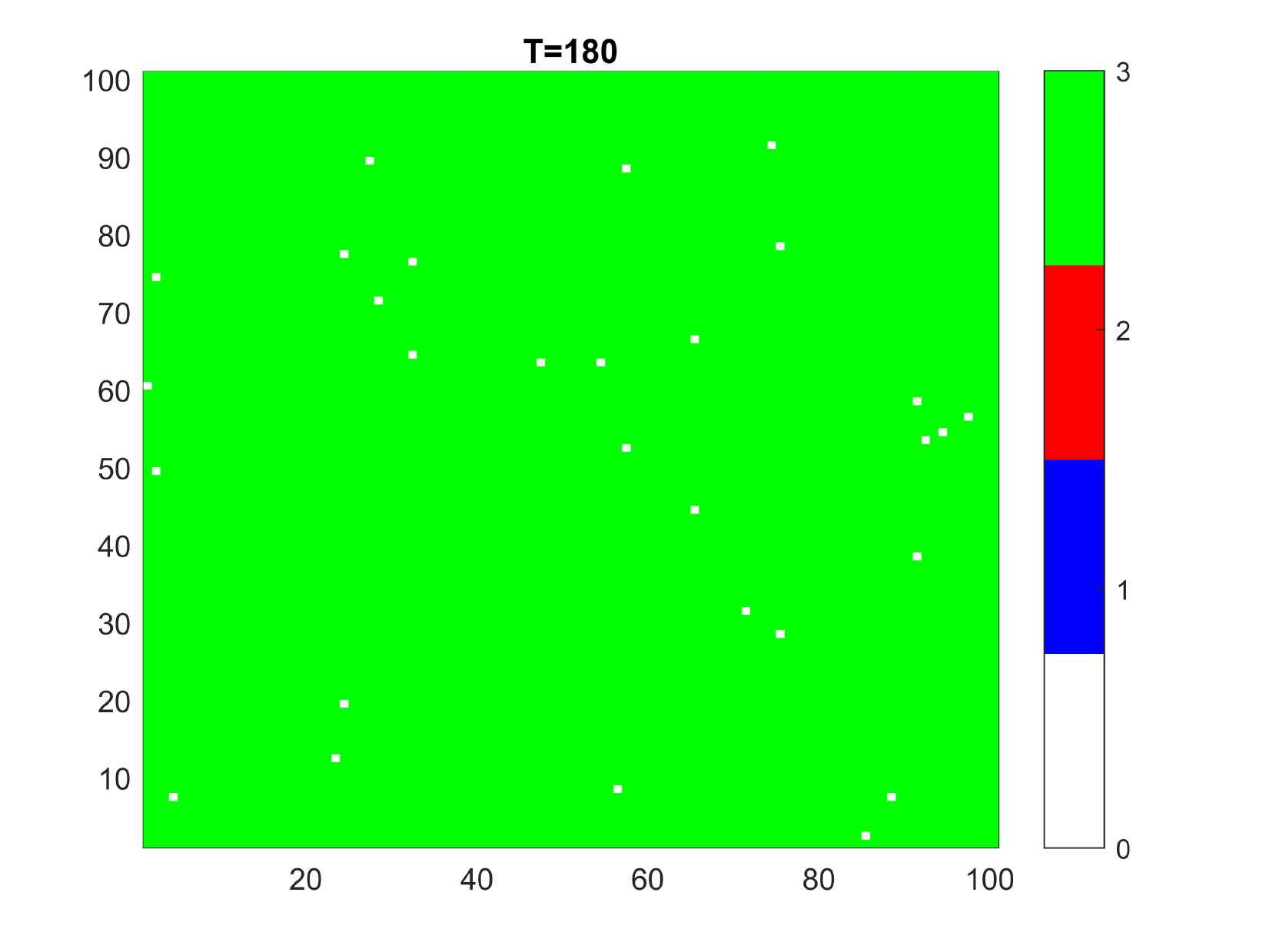}
		\end{tabular}
		\caption{{Plots of the temporal and spatial behavior of the disease spread for $n=1$.}\label{n_1_plots}}
	\end{figure}
	In Fig.~\ref{n_1_plots}, the first plot shows the temporal behavior of the exposed cases ($E(t)$) and the infectious cases ($I(t)$) of the epidemic. These temporal plots are averaged on 50 simulation samples. The rest plots of Fig.~\ref{n_1_plots} are CA plots that represent the spatial evolution of disease spread. From these CA plots, we can hardly detect any clustering of the infected cases. This happens because the average interaction distance, $\langle d\rangle \approx 11.11$. Thus a susceptible person can be infected by an infectious person who is far away from the susceptible one.

	\begin{figure}[H]
		\begin{tabular}{ccc}
			\centering
			\includegraphics[scale=0.22]{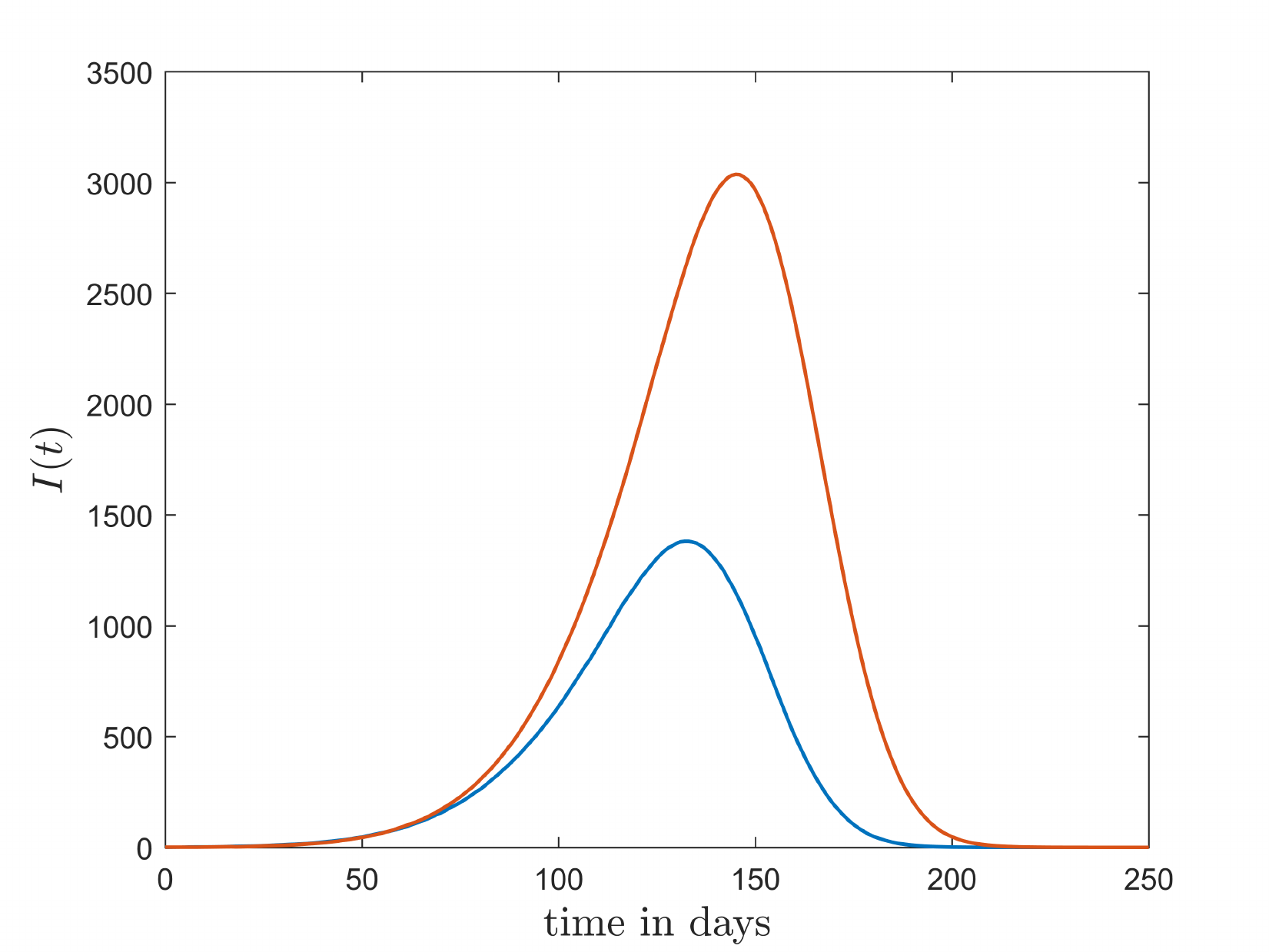}&
			\includegraphics[scale=0.22]{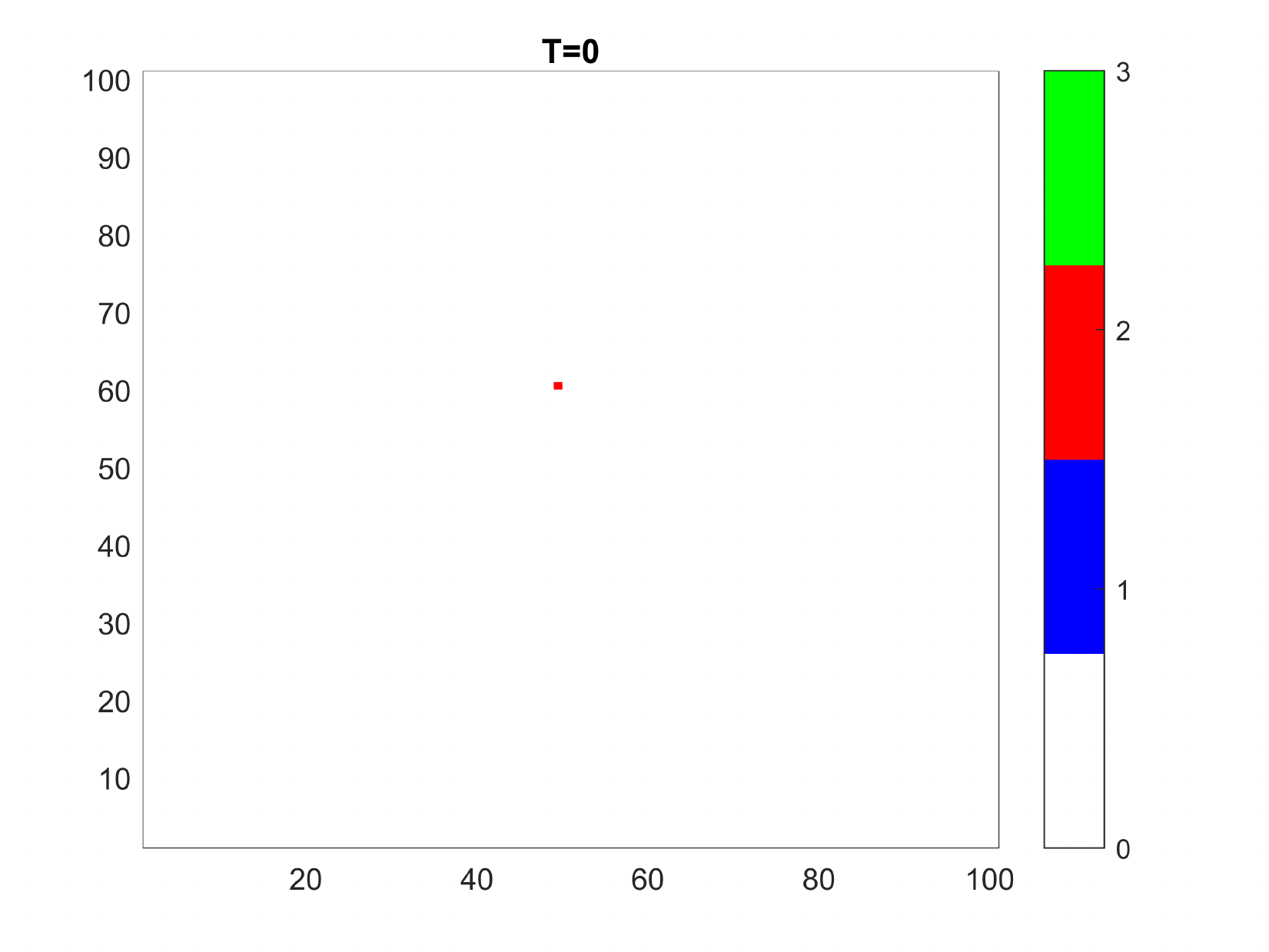}&
			\includegraphics[scale=0.22]{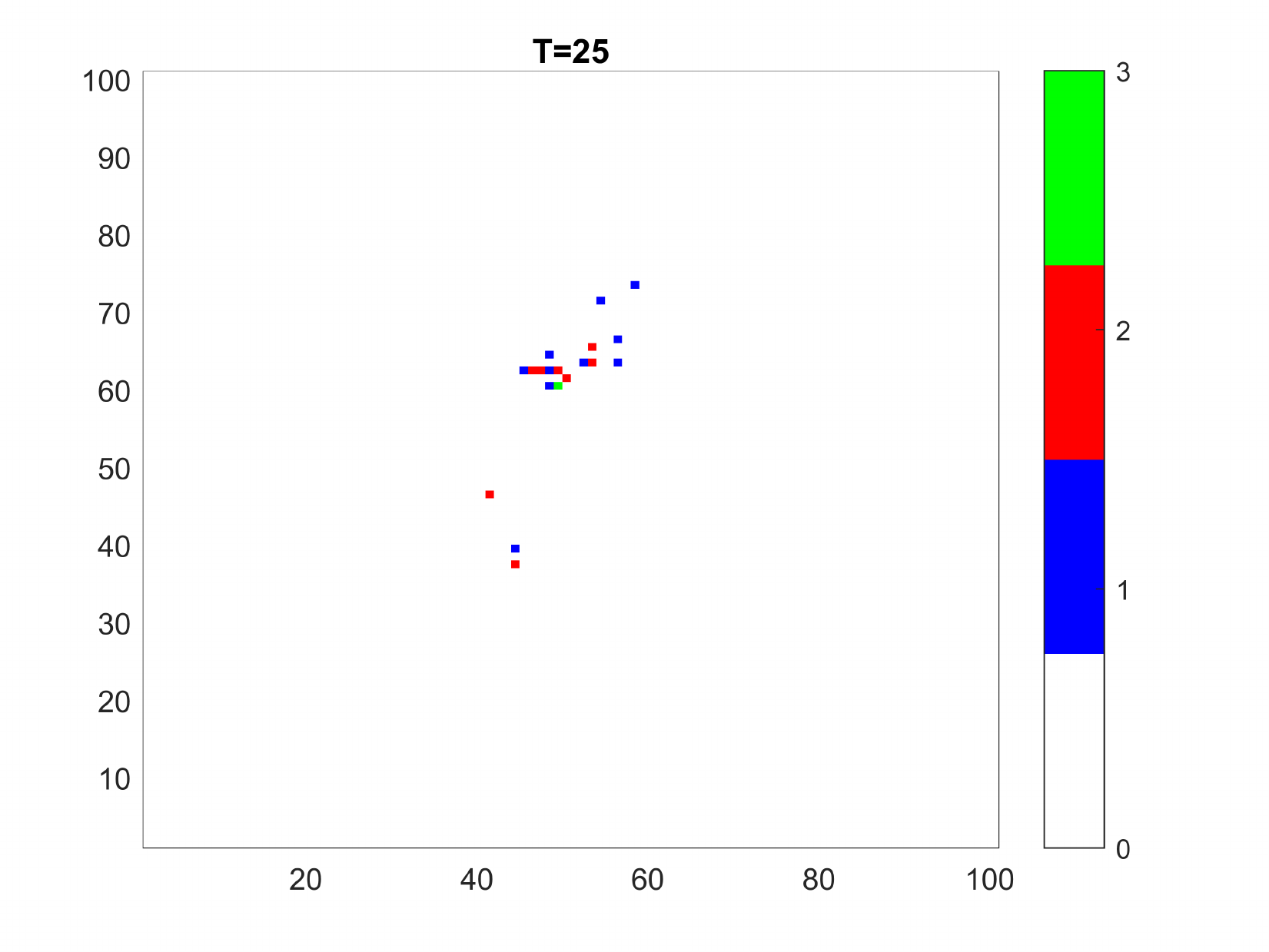}\\
			\includegraphics[scale=0.22]{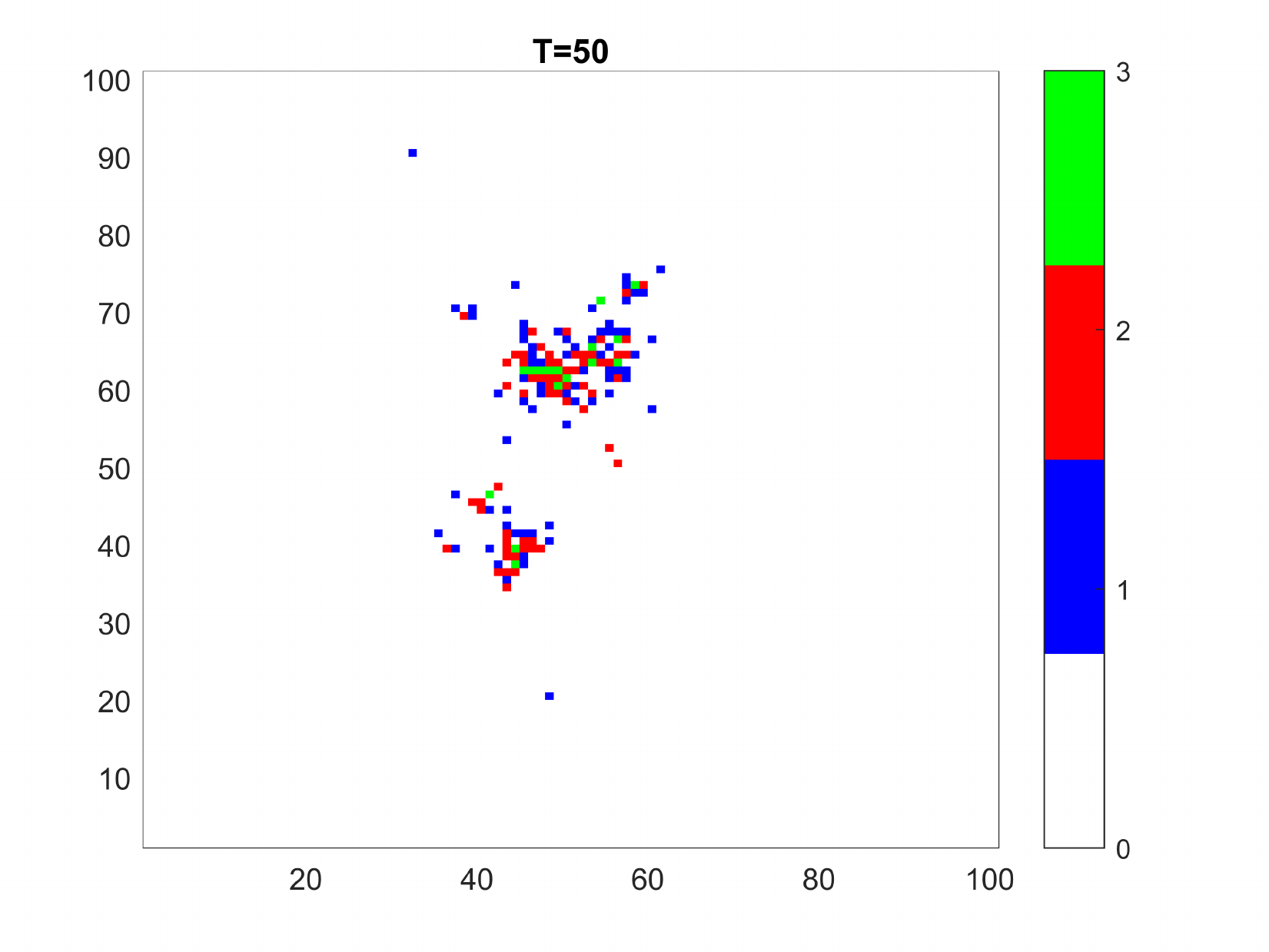}&
			\includegraphics[scale=0.22]{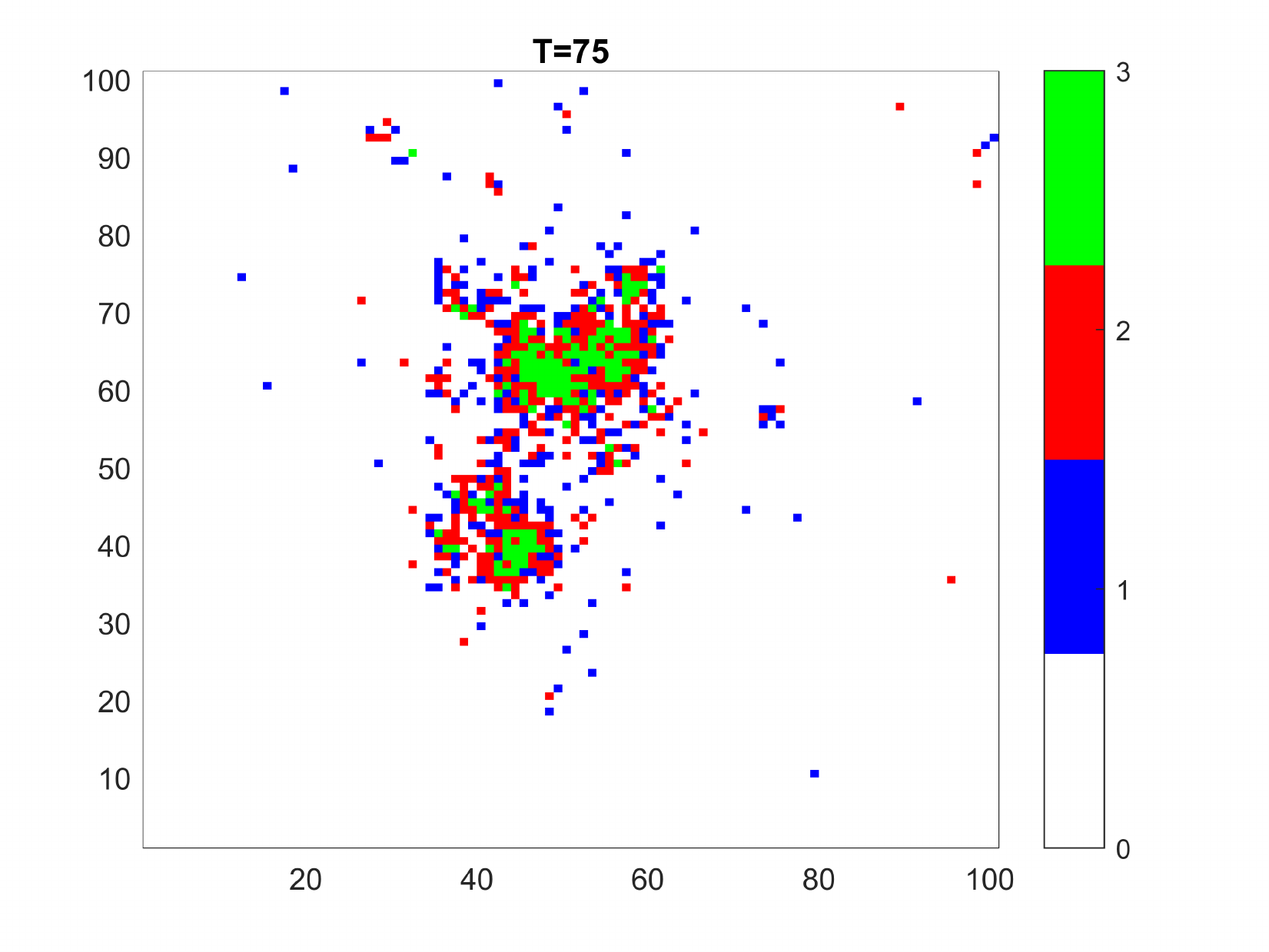}&
			\includegraphics[scale=0.22]{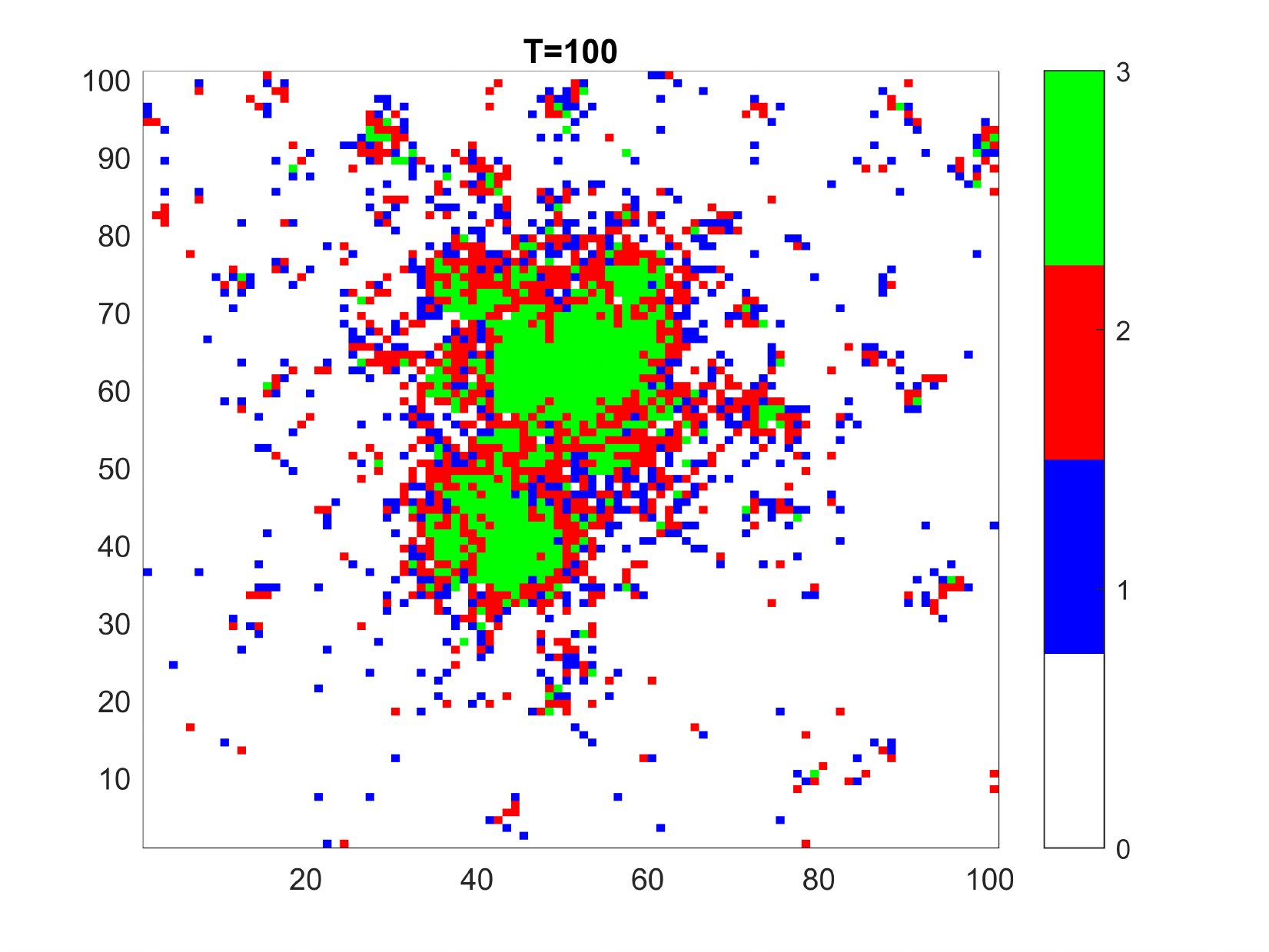}\\
			\includegraphics[scale=0.22]{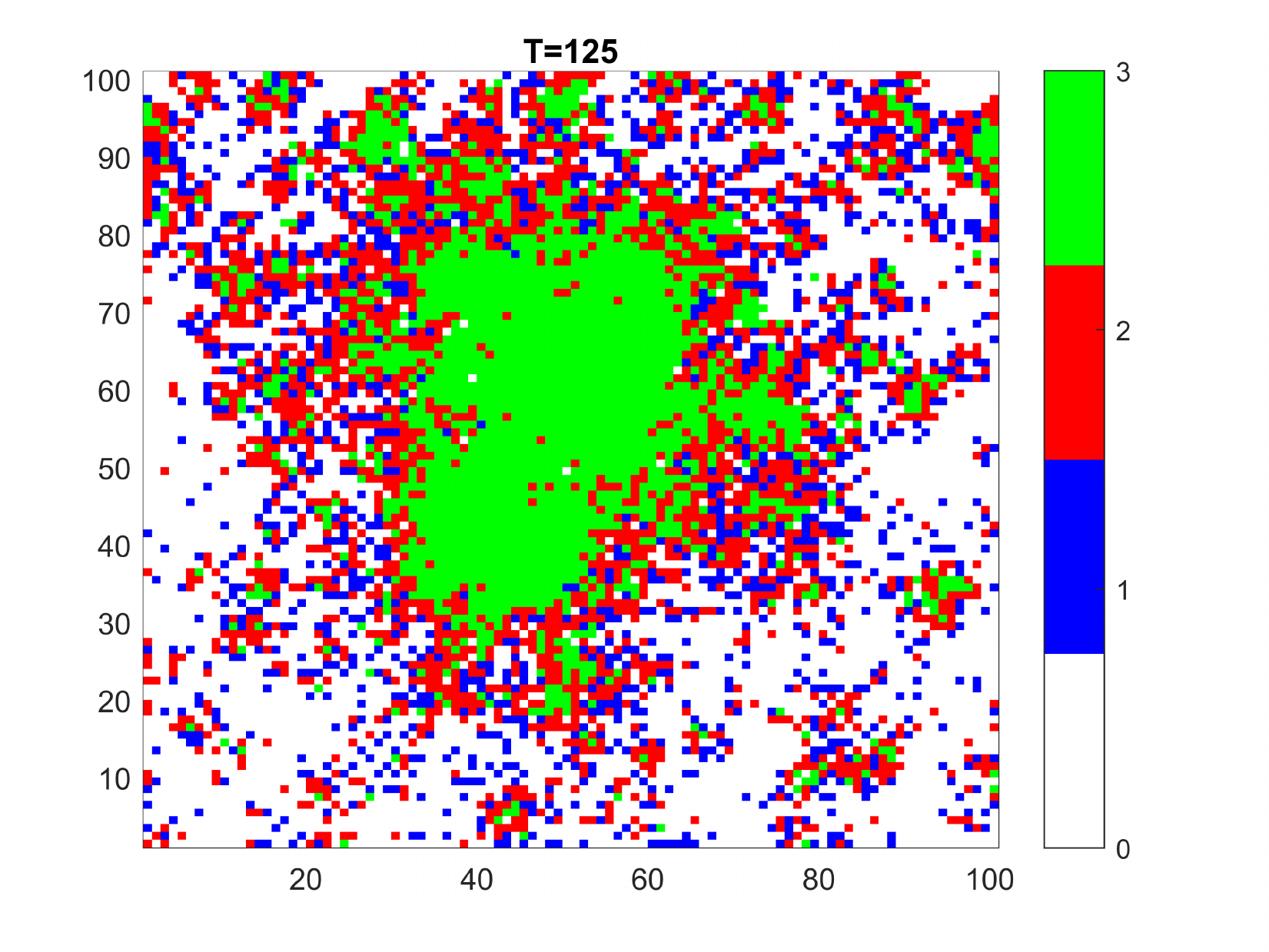}&
			\includegraphics[scale=0.22]{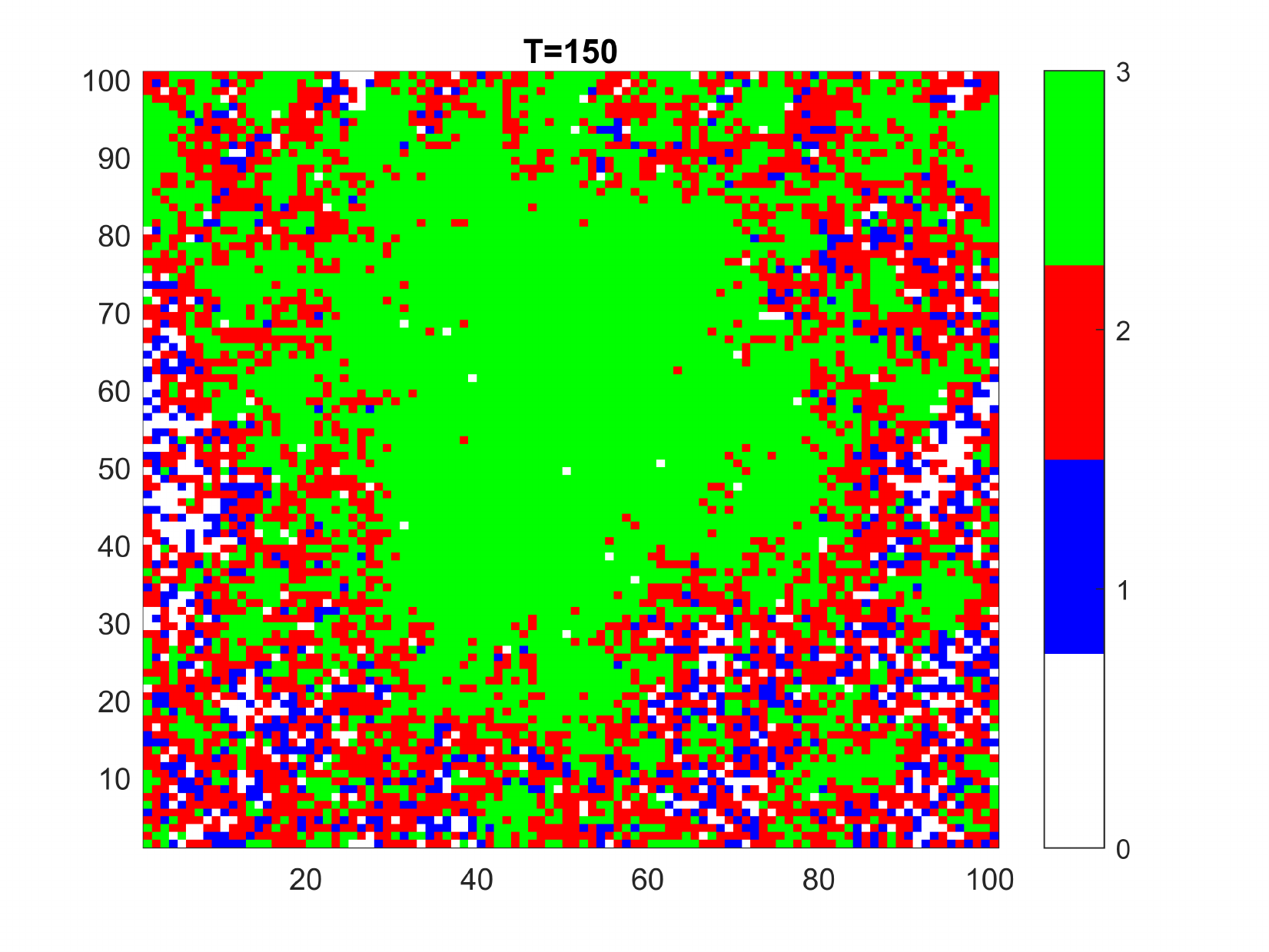}&
			\includegraphics[scale=0.22]{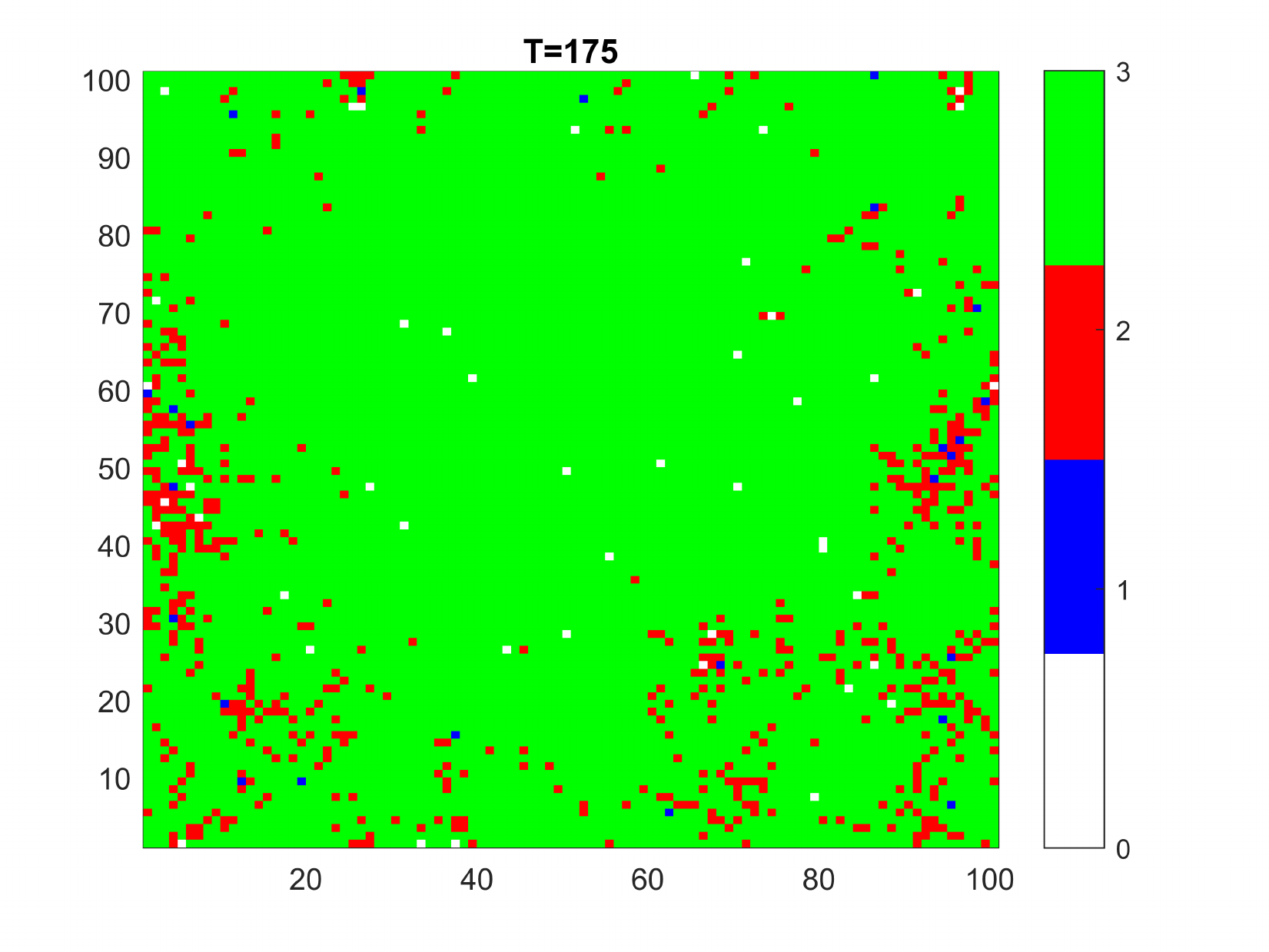}\\
			\includegraphics[scale=0.22]{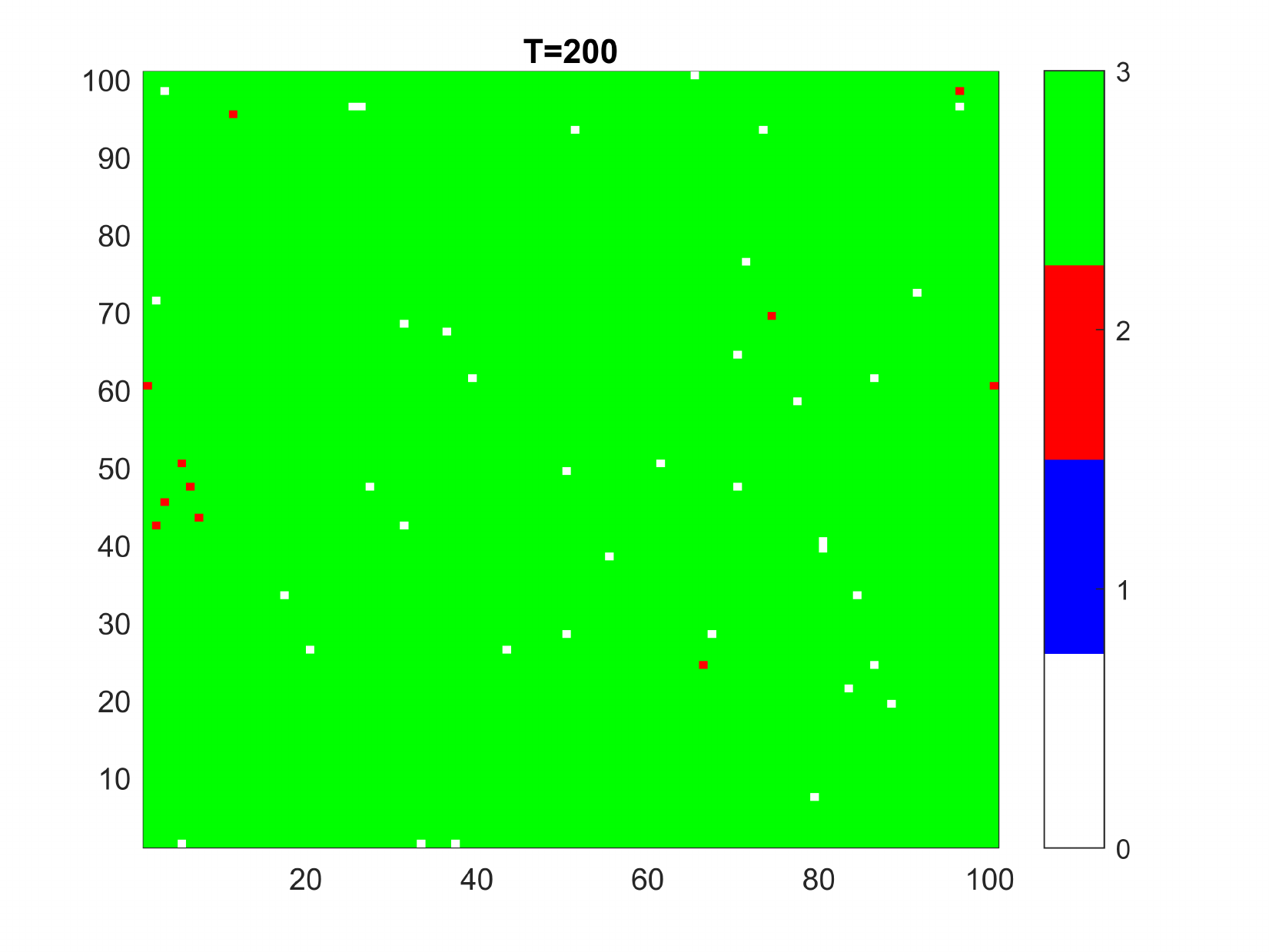}&
			\includegraphics[scale=0.22]{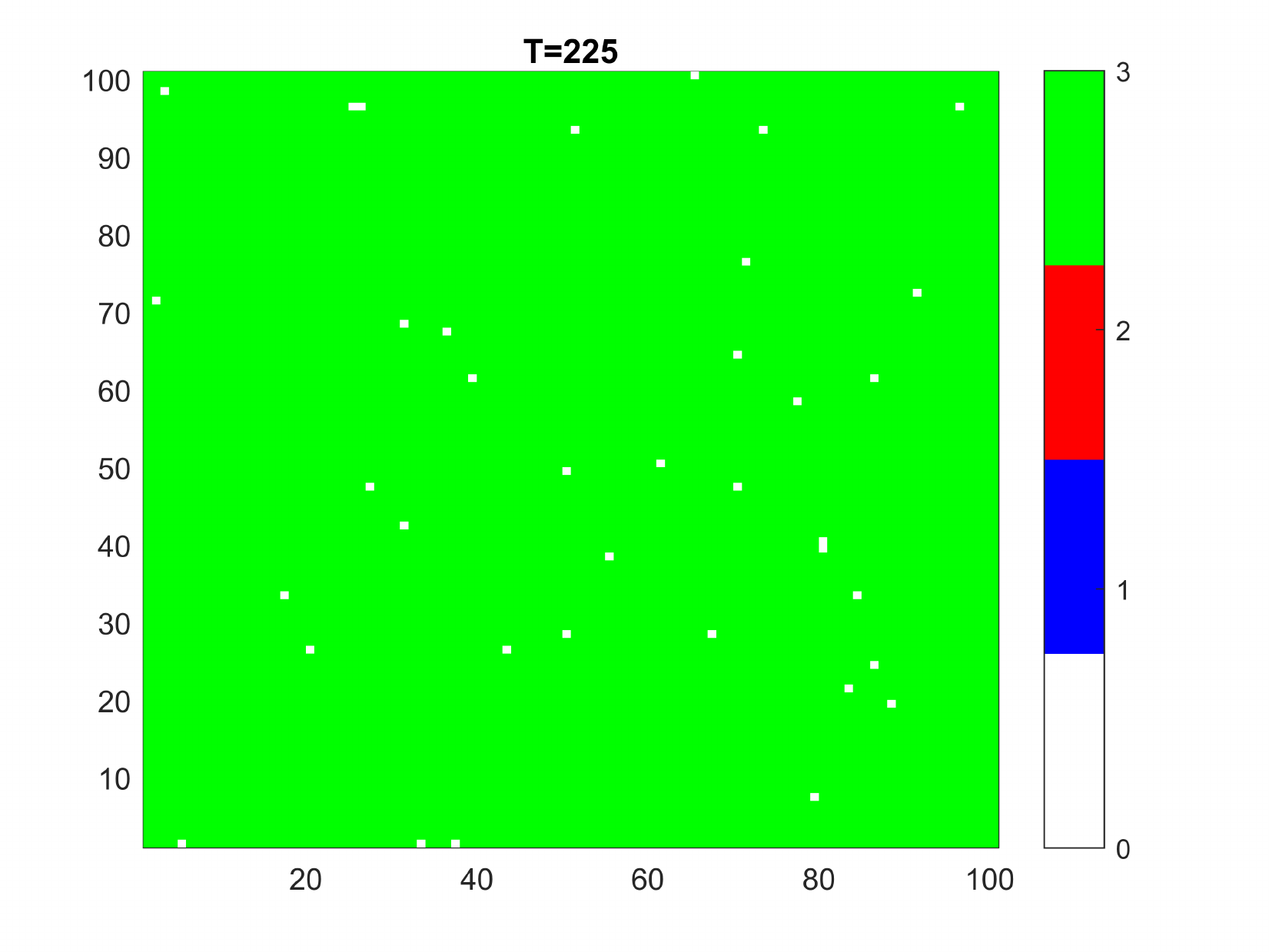}&
			\includegraphics[scale=0.22]{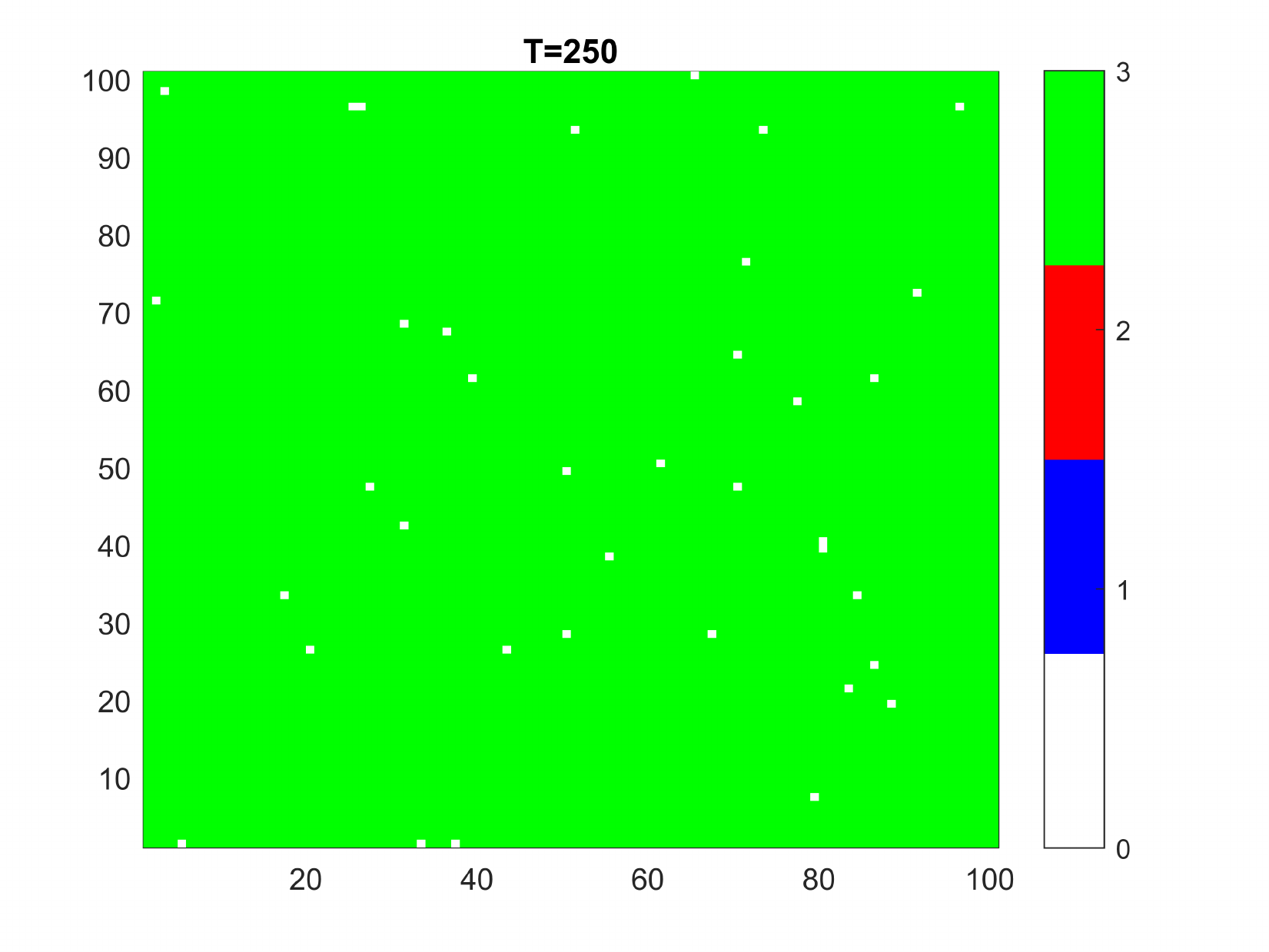}
		\end{tabular}
		\caption{{Plots of the temporal and spatial behavior of the disease spread for $n=2$.}\label{n_2_plots}}
	\end{figure}
	In Fig.~\ref{n_2_plots}, the first plot again shows the temporal growth of the epidemic. It can be seen from the CA plots that for $n=2$, clusters are formed. The reason behind this is the short average interaction distance. For $n=2$, the average interaction distance, $\langle d\rangle \approx 2.77$. Also from the temporal plot, we can see that the infection spread time is increased than in the $n=1$ case. This is also because of the short average interaction distance. For a short average interaction distance, only a few susceptible persons can interact with the infectious person. So, if most of those susceptible persons become infected then the infectious person cannot spread the disease further. Whereas, for $n=1$ case, an infectious person can interact with many susceptible persons. Hence, an infectious person can infect more people during the infectious period.
	
	\begin{figure}[H]
		\begin{tabular}{ccc}
			\centering
			\includegraphics[scale=0.22]{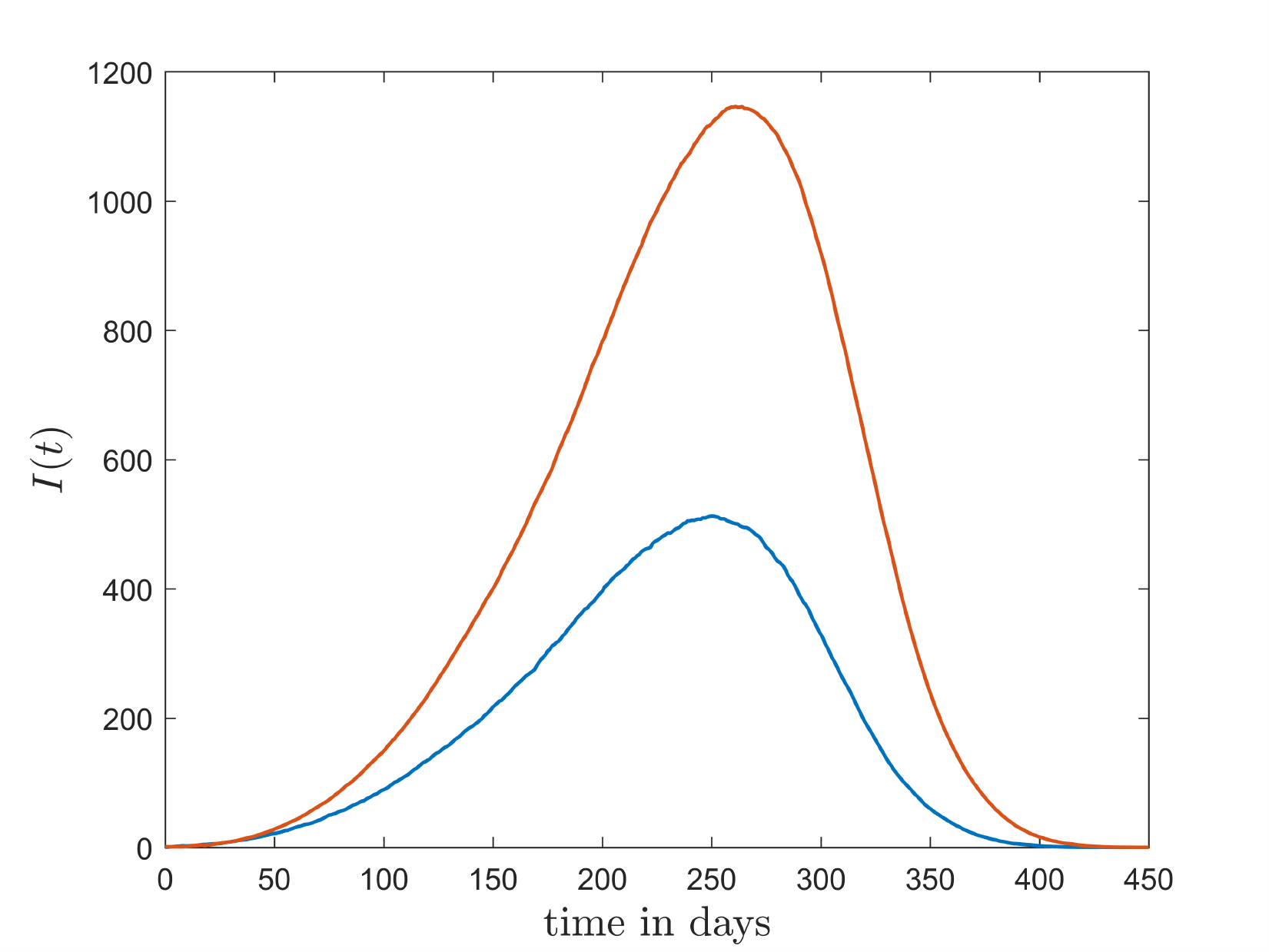}&
			\includegraphics[scale=0.22]{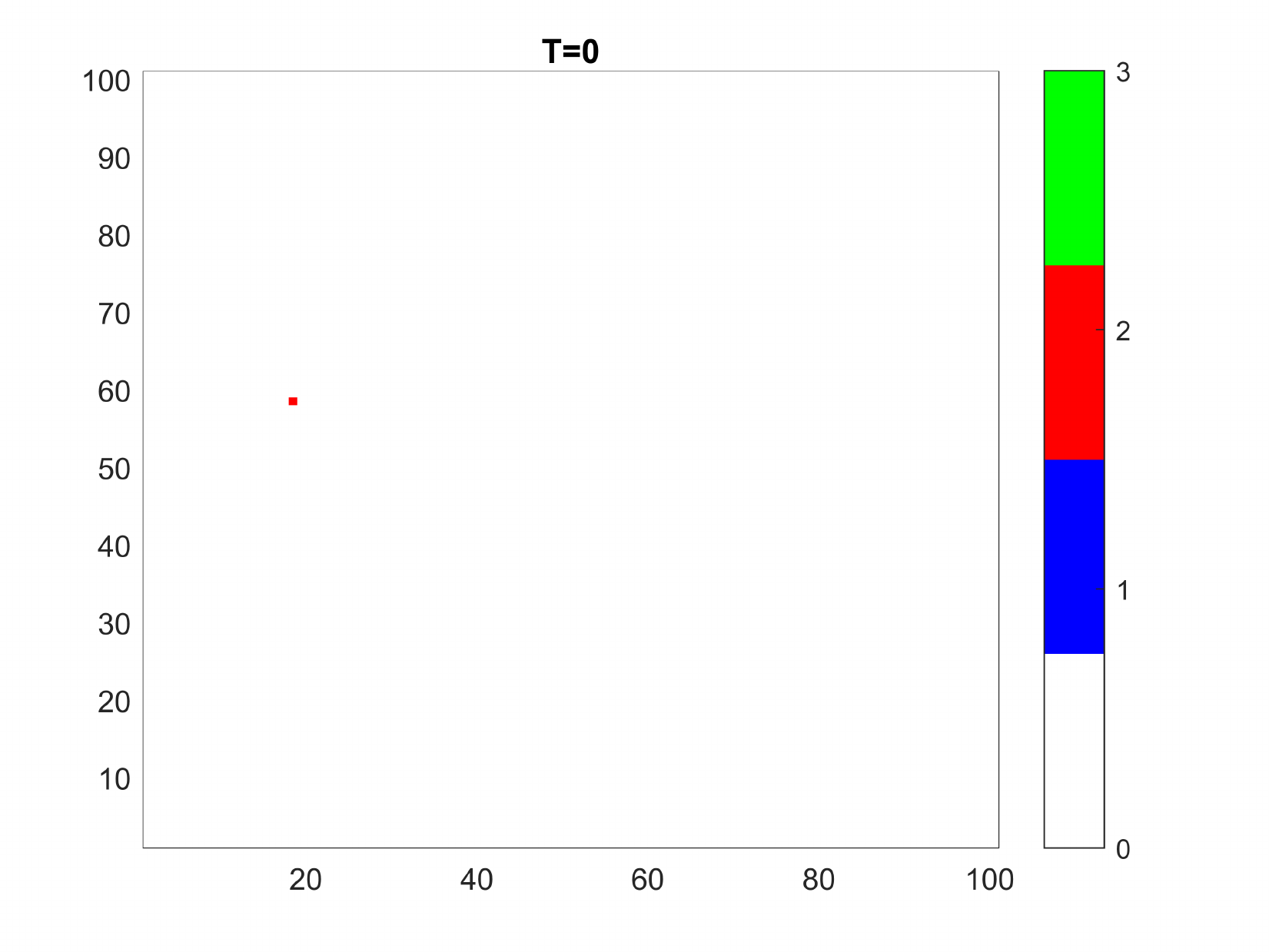}&
			\includegraphics[scale=0.22]{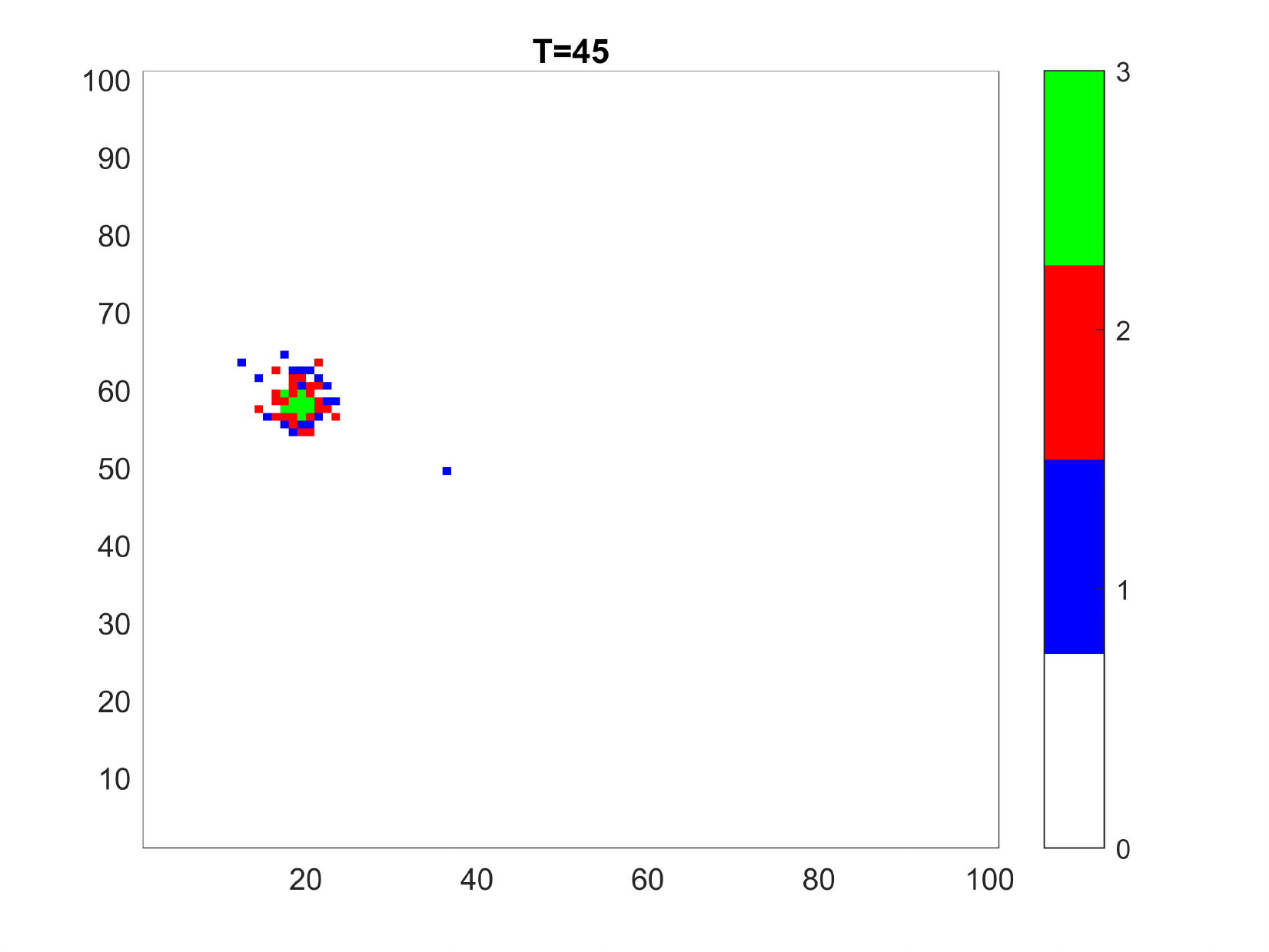}\\
			\includegraphics[scale=0.22]{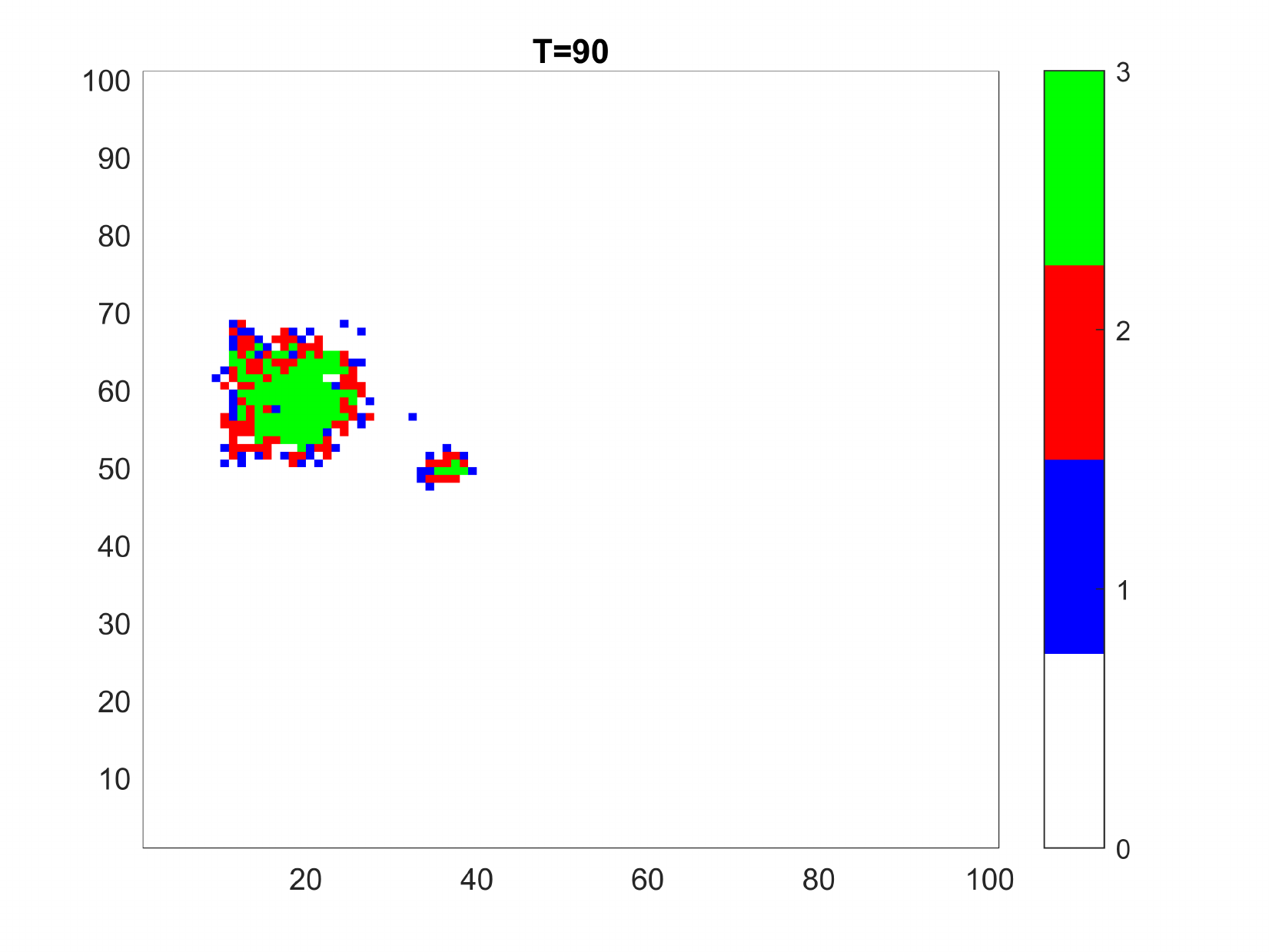}&
			\includegraphics[scale=0.22]{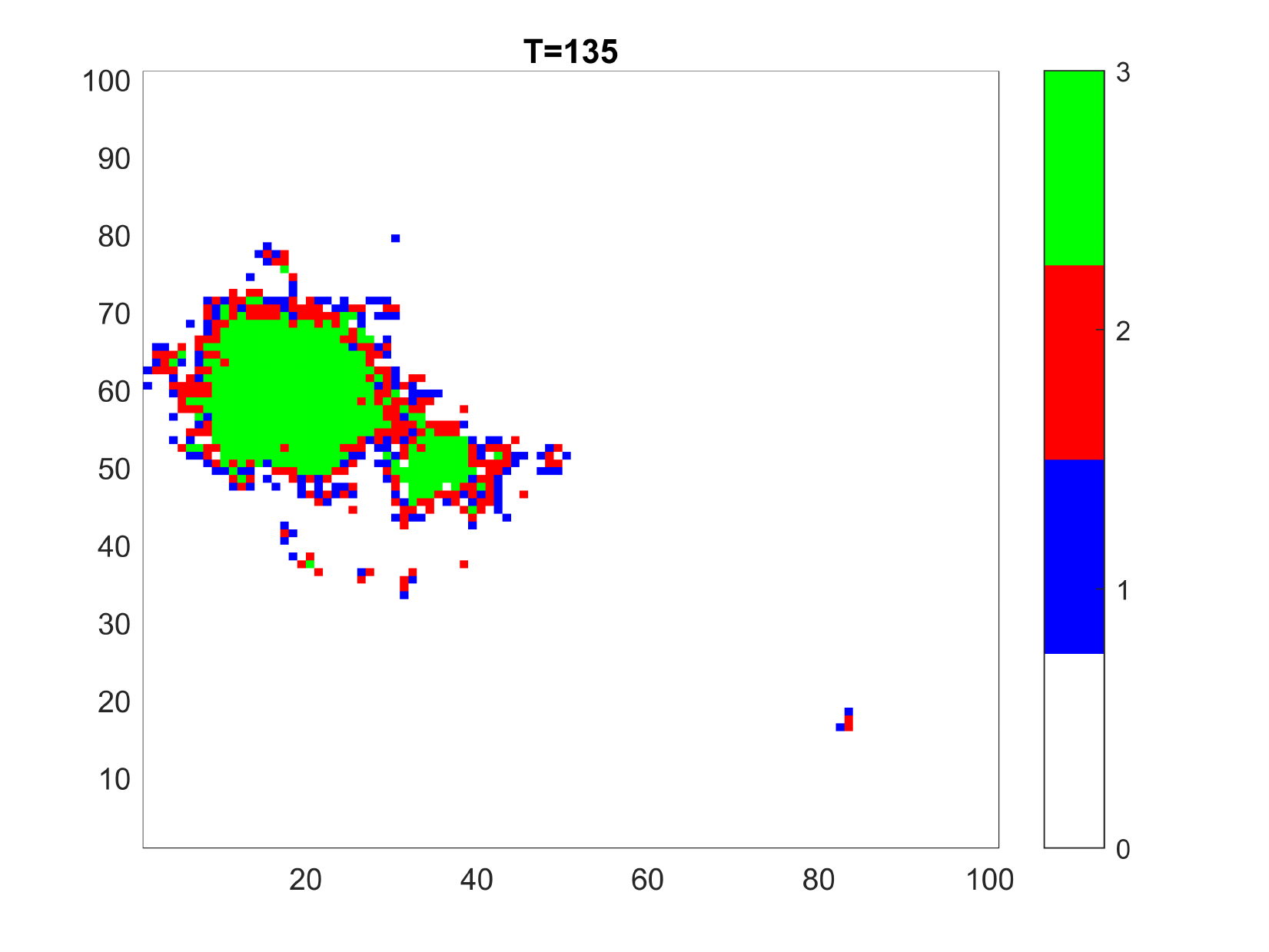}&
			\includegraphics[scale=0.22]{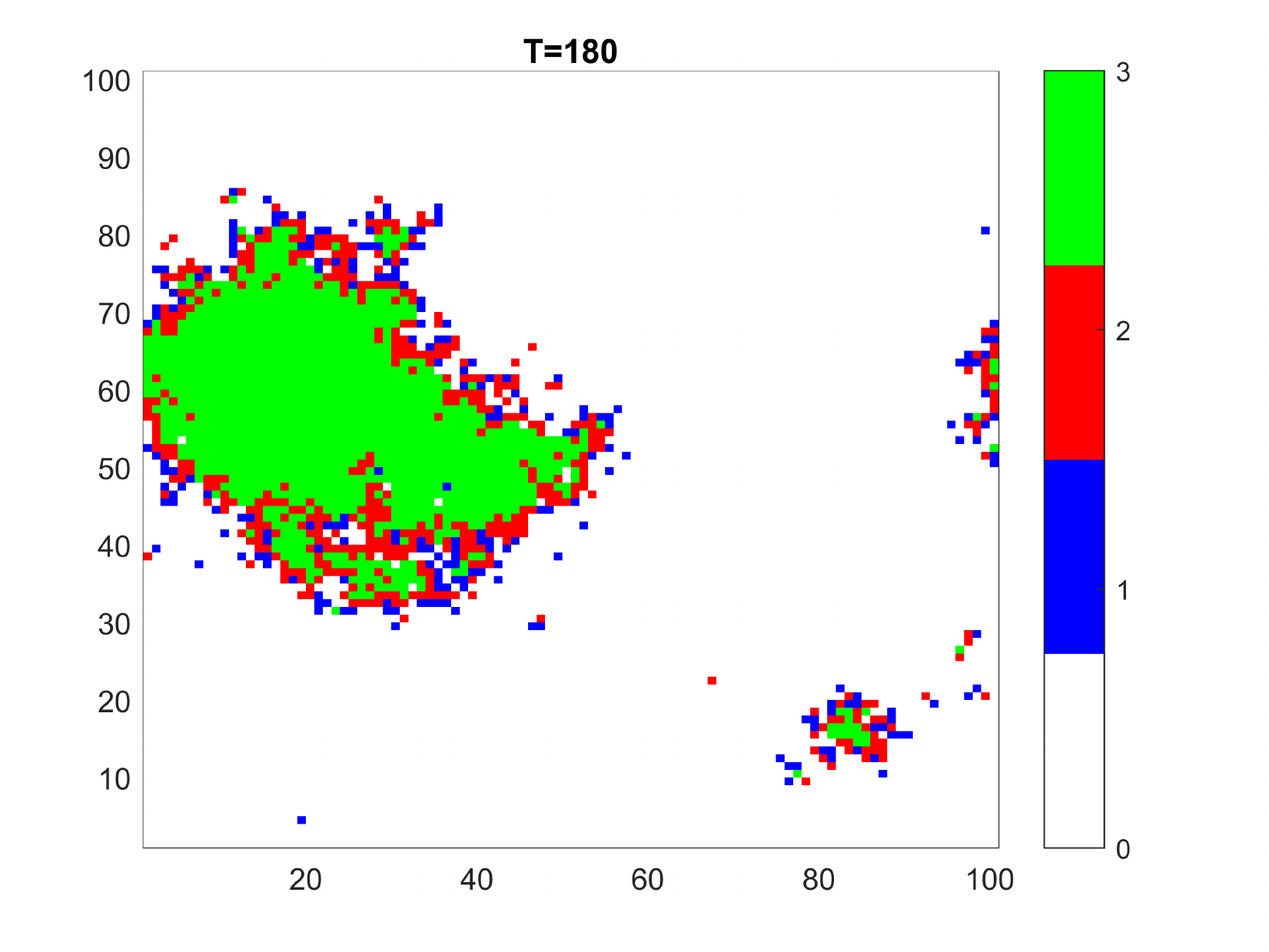}\\
			\includegraphics[scale=0.22]{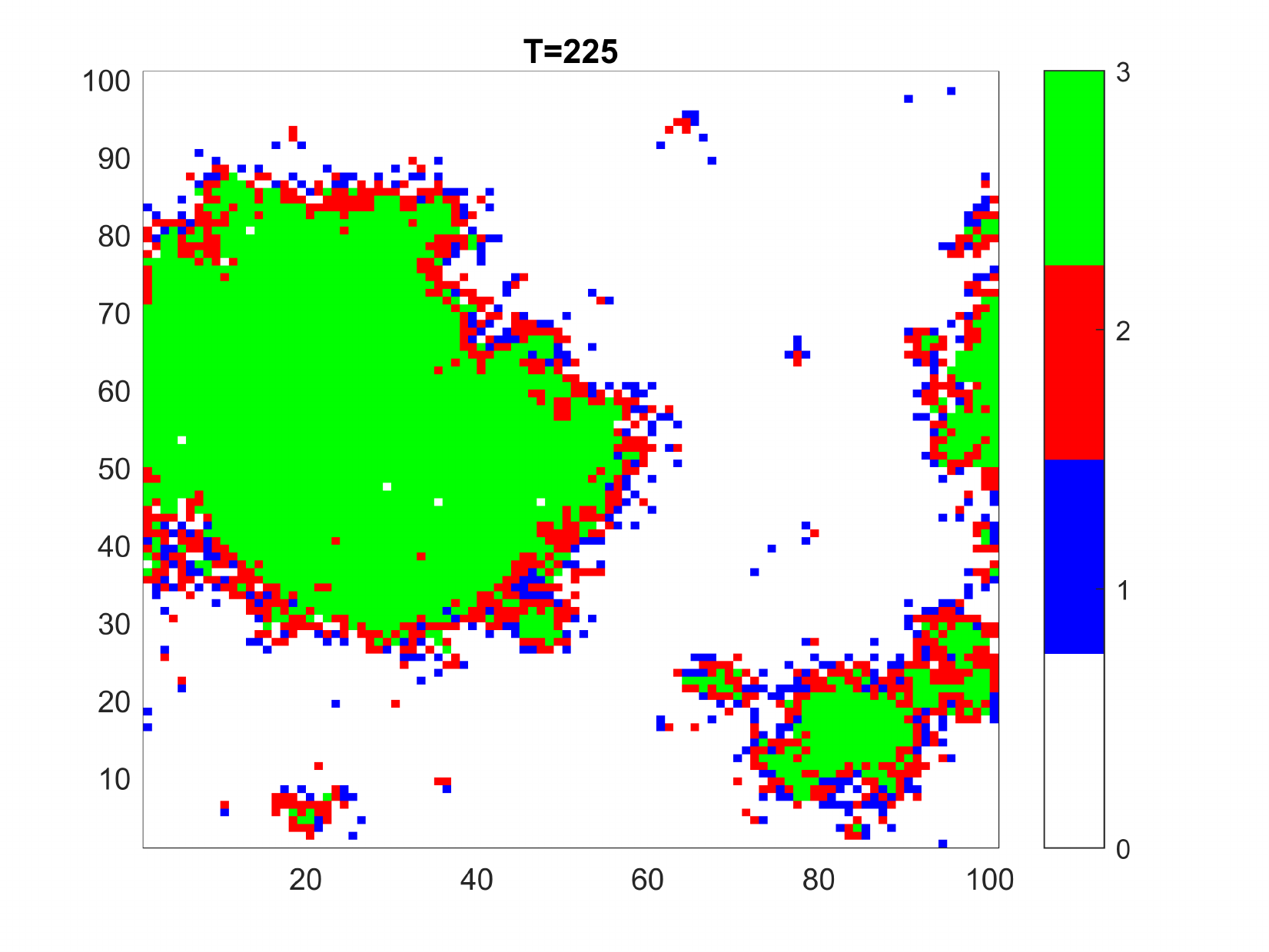}&
			\includegraphics[scale=0.22]{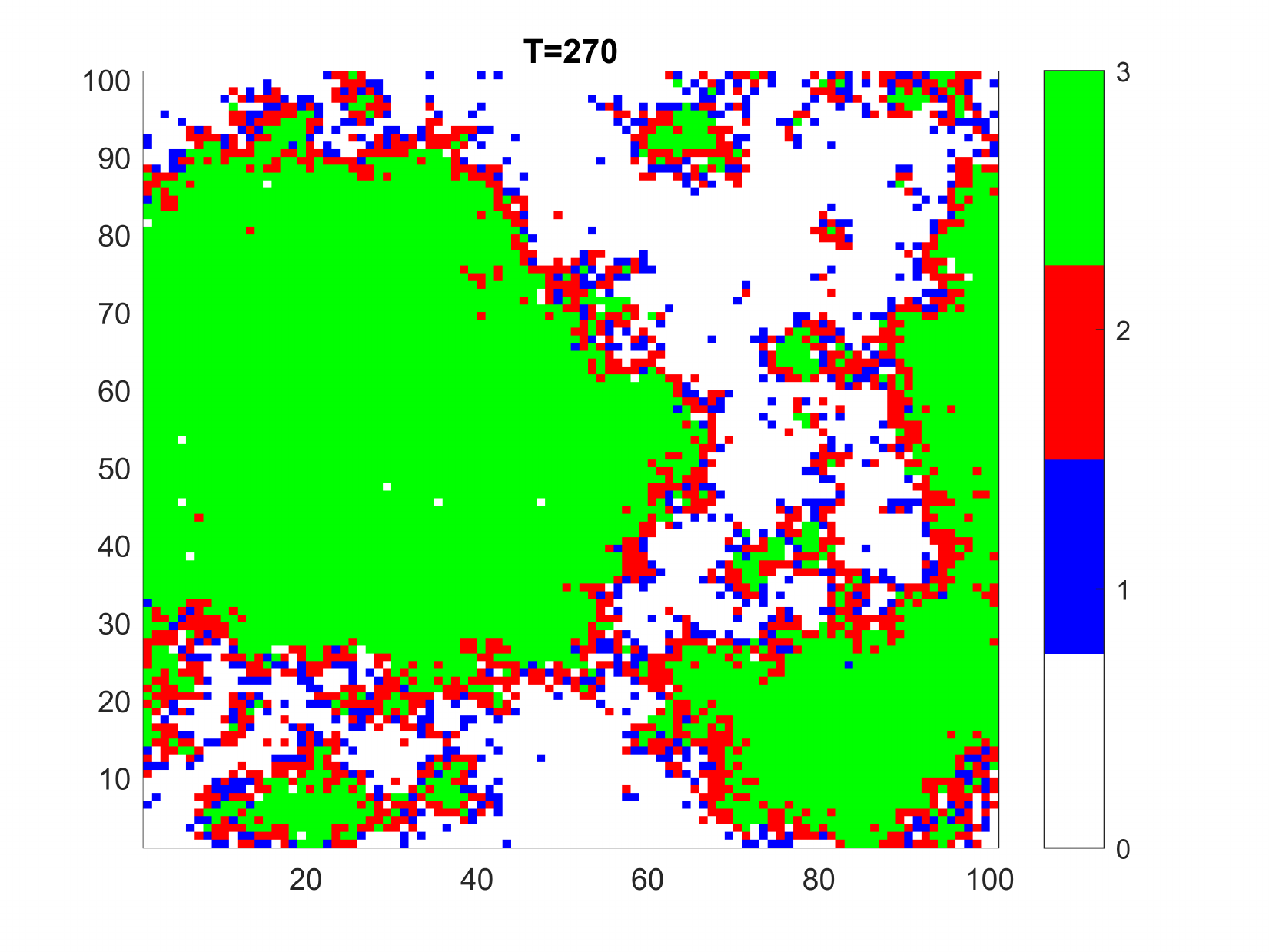}&
			\includegraphics[scale=0.22]{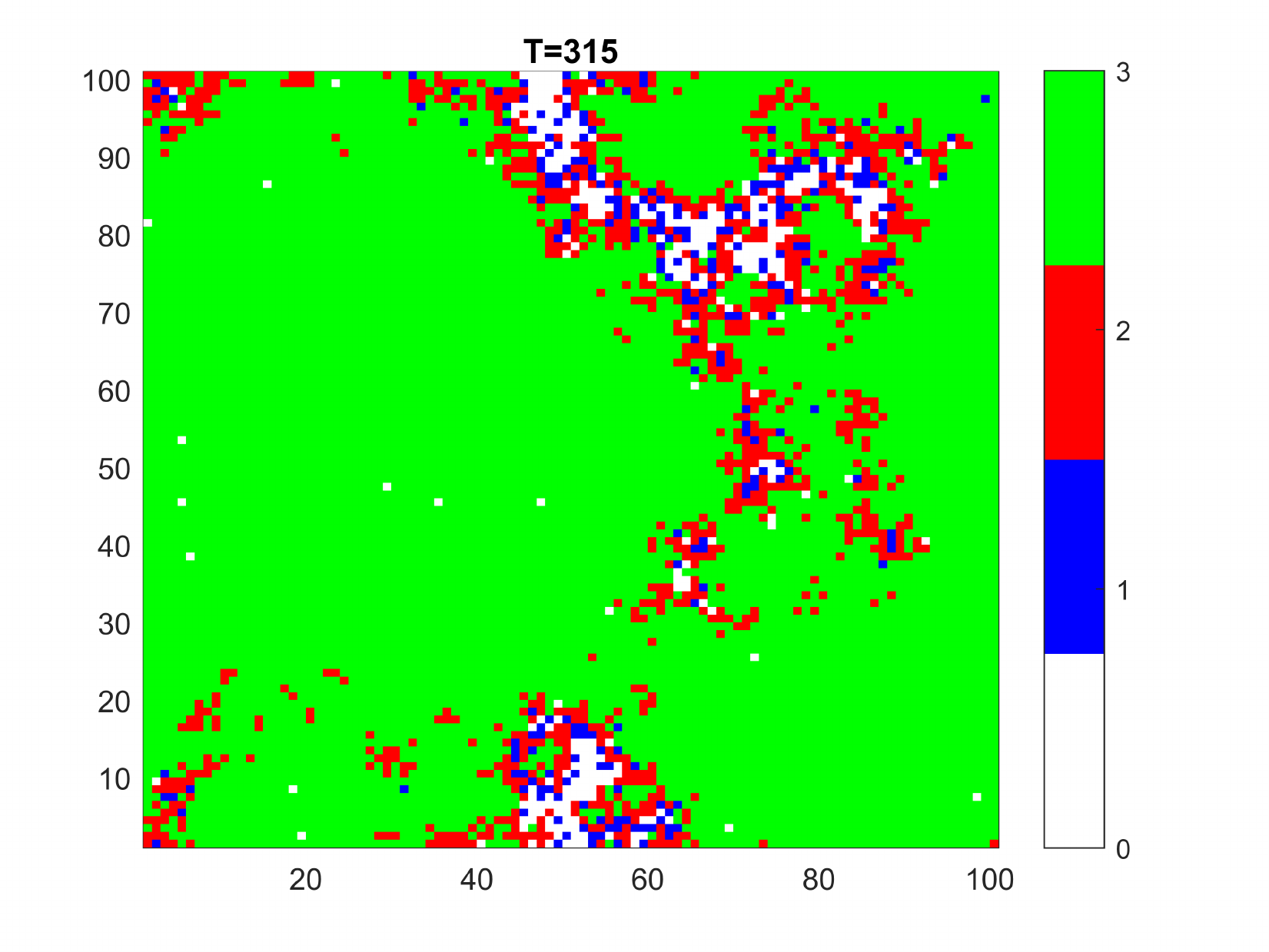}\\
			\includegraphics[scale=0.22]{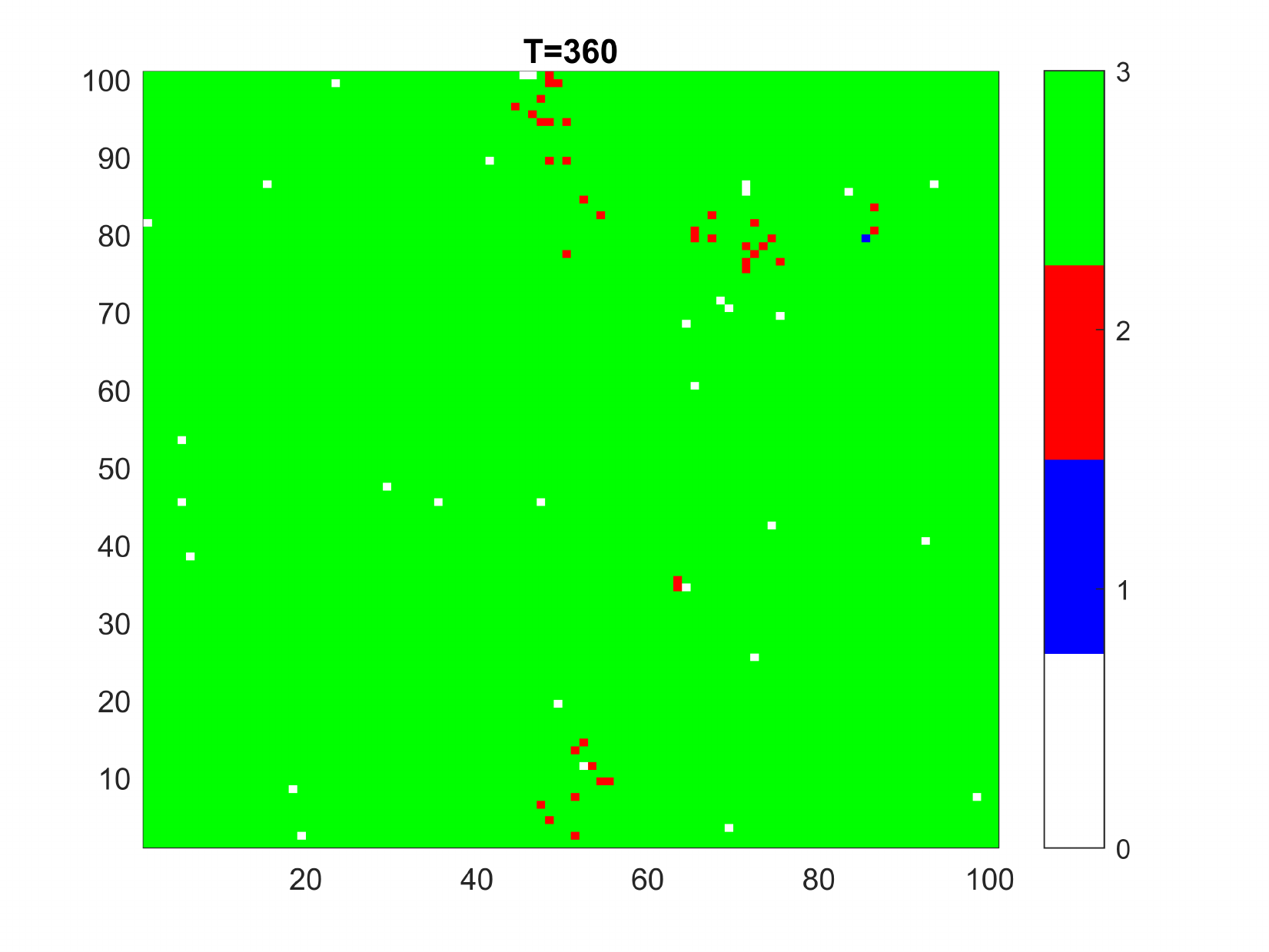}&
			\includegraphics[scale=0.22]{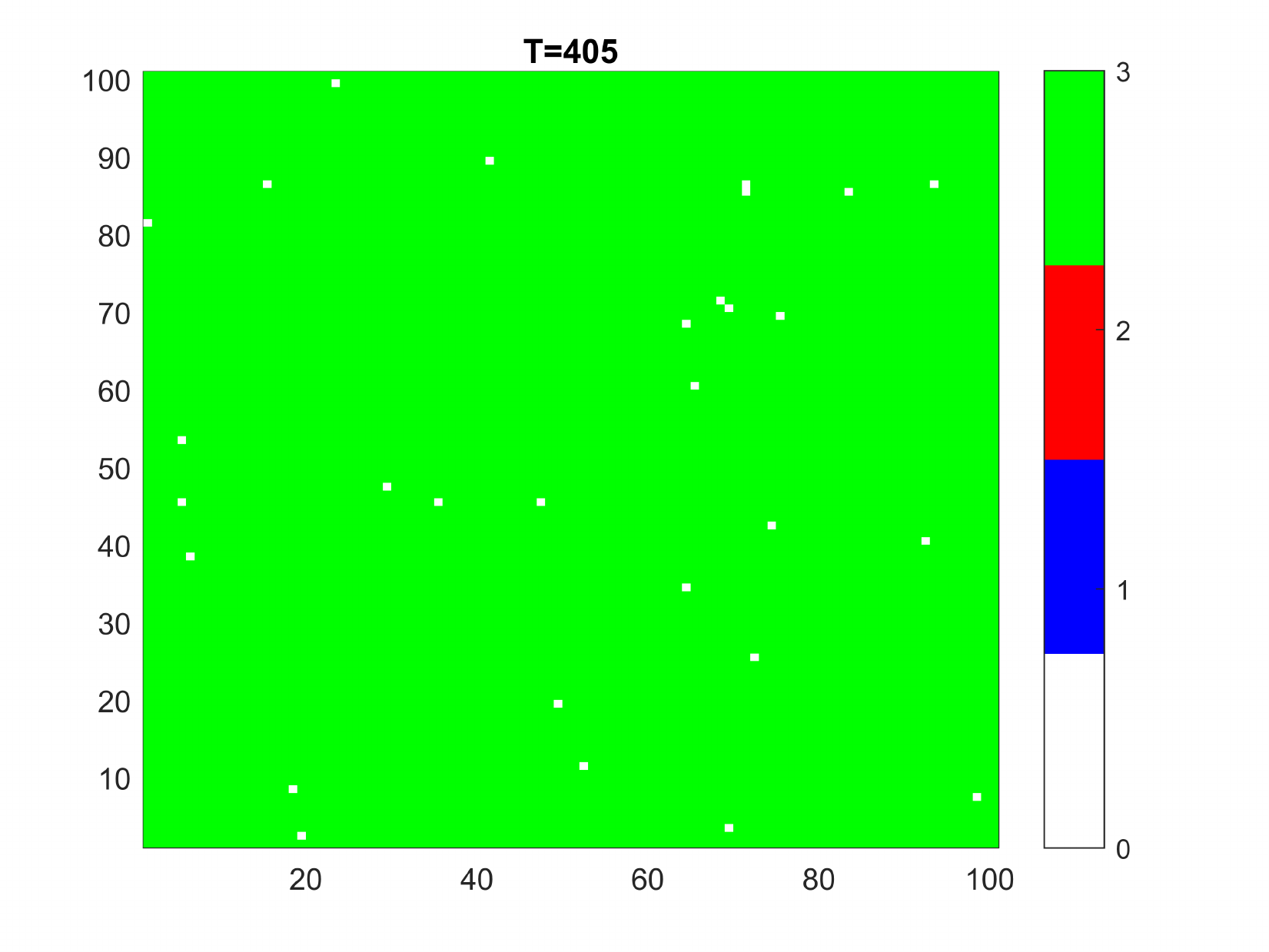}&
			\includegraphics[scale=0.22]{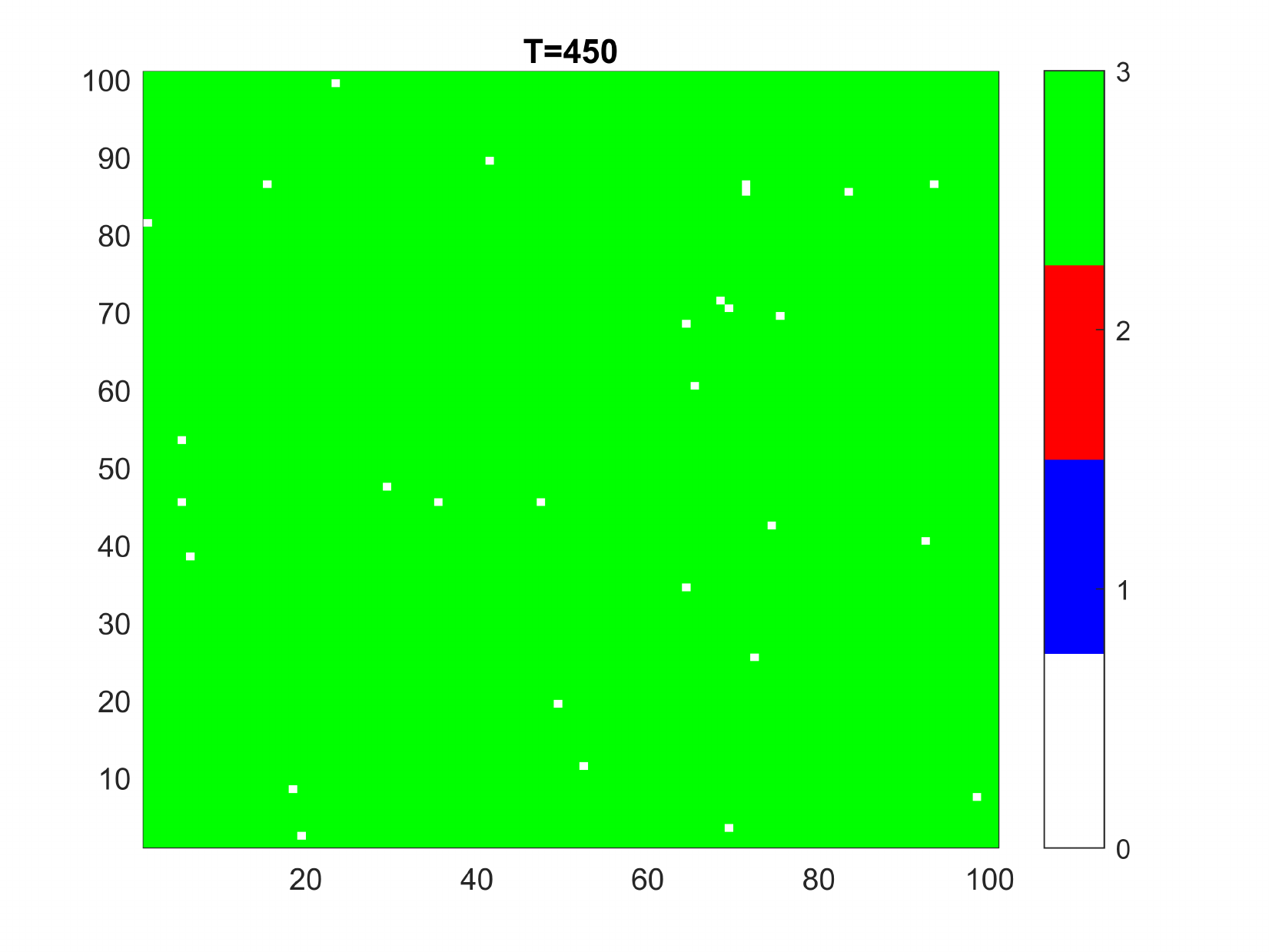}
		\end{tabular}
		\caption{{Plots of the temporal and spatial behavior of the disease spread for $n=3$.}\label{n_3_plots}}
	\end{figure}
	Fig.~\ref{n_3_plots} shows the evolution of the disease for $n=3$. Here we can also clearly find the clusters. These clusters are more prominent than the $n=2$ case because of the less average interaction distance ($\langle d\rangle$). Value of the average interaction distance for $n=3$ is $\langle d\rangle\approx 1.35$. Here we can see that the disease takes a longer time to fall for $n=3$ than for $n=1$ and $n=2$. The reason behind this is the lower value of the average interaction distance which is discussed earlier.
	
	Hence, from the above discussions, we can conclude that the clustering behavior of the disease spread depends on the average interaction distance ($\langle d\rangle$) as well as on degree exponent $n$. Also, the average interaction distance ($\langle d\rangle$), gives an average estimation of the number of susceptible persons who can interact with an infectious person which is represented by $8\langle d\rangle$. So, for a large  $\langle d\rangle$ (or small $n$) an infectious person can spread the disease to distanced region. Thus the infection period depends on the average interaction distance ($\langle d\rangle$) and also on $n$.
	
\section{Comparison with data}\label{data_analysis}
	In this section, we have tried to fit our model with current COVID-19 data. Our model has four free parameters which are, (i)  $q$ : disease transmission probability, (ii) $n$ : degree exponent, (iii) $\tau_{I}$ : mean latency period, (iv) $\tau_{R}$ : mean infectious period. We have optimized these free parameters for different waves of the COVID-19 pandemic in India. Here, we have considered each wave separately and normalized the active cases of each wave with the total number of infected cases in the respective wave. The data is taken from covid19india.org \cite{data}. The date range of the different waves that we have considered here are given below:
	\begin{table}[H]
		\centering
		\begin{tabular}{|c|c|c|}
			\hline
			Waves & Start date & End date\\
			\hline
			First wave& 30-Jan-2020 & 16-Feb-2021\\
			Second wave & 17-Feb-2021 & 31-Oct-2021\\
			\hline                     
		\end{tabular}
		\caption{{Date ranges for different waves.} \label{drange_tab}}
	\end{table}
	
	To fit the model with the data we have optimized the sum of squared errors ($SSE$)
	
	\begin{equation}
		SSE=\sum_{k}\left(i_{k}^{d}-i_{k}\right)^{2} \label{SSE}
	\end{equation}	
	$i_{k}^{d}$ : fraction of the active cases from the data.
	$i_{k}$ : fraction of the infectious cases from the model.
	 
	 The results of the best fit parameter values are given below.
 	\begin{table}[H]
	 	\centering
	 	\begin{tabular}{|c|c|c|c|c|}
	 		\hline
	 		Waves & q & n & $\tau_{I}~\left(days\right)$ & $\tau_{R}~\left(days\right)$\\
	 		\hline
	 		First wave& 0.1950 & 1.9310 & 5 & 11\\
	 		Second wave & 0.2406 & 1.3449 & 8 & 10\\
	 		\hline                     
	 	\end{tabular}
	 	\caption{{Fitting parameter values for different waves.} \label{fitval_tab}}
	 \end{table}
 
\begin{figure}[H]
	 	\begin{subfigure}{.5\textwidth}
	 		\centering
	 		\includegraphics[scale=0.35]{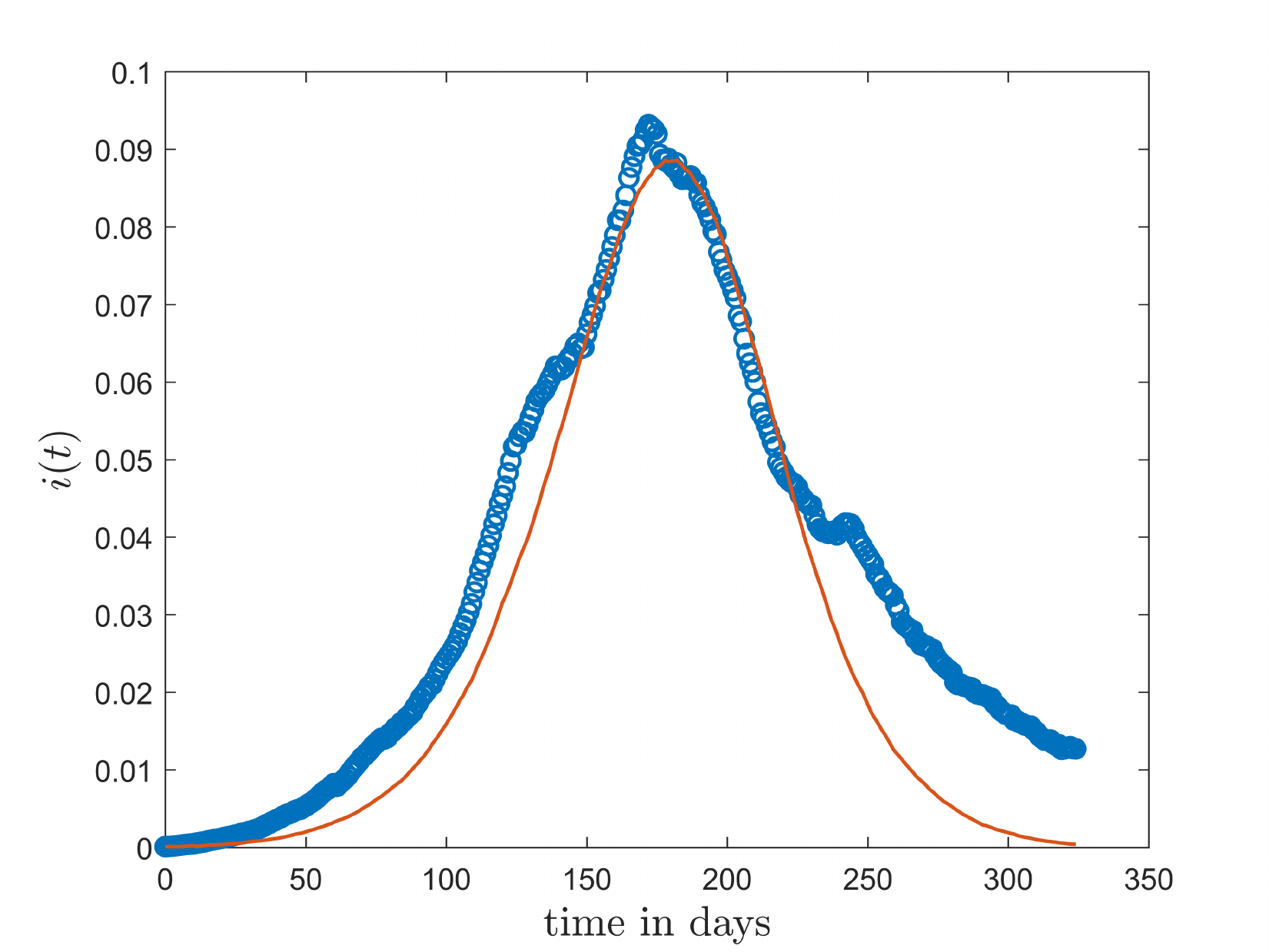}
	 		\caption{}
	 		\label{F_wave}
	 	\end{subfigure}
	 	\begin{subfigure}{.5\textwidth}
	 		\centering
	 		\includegraphics[scale=0.35]{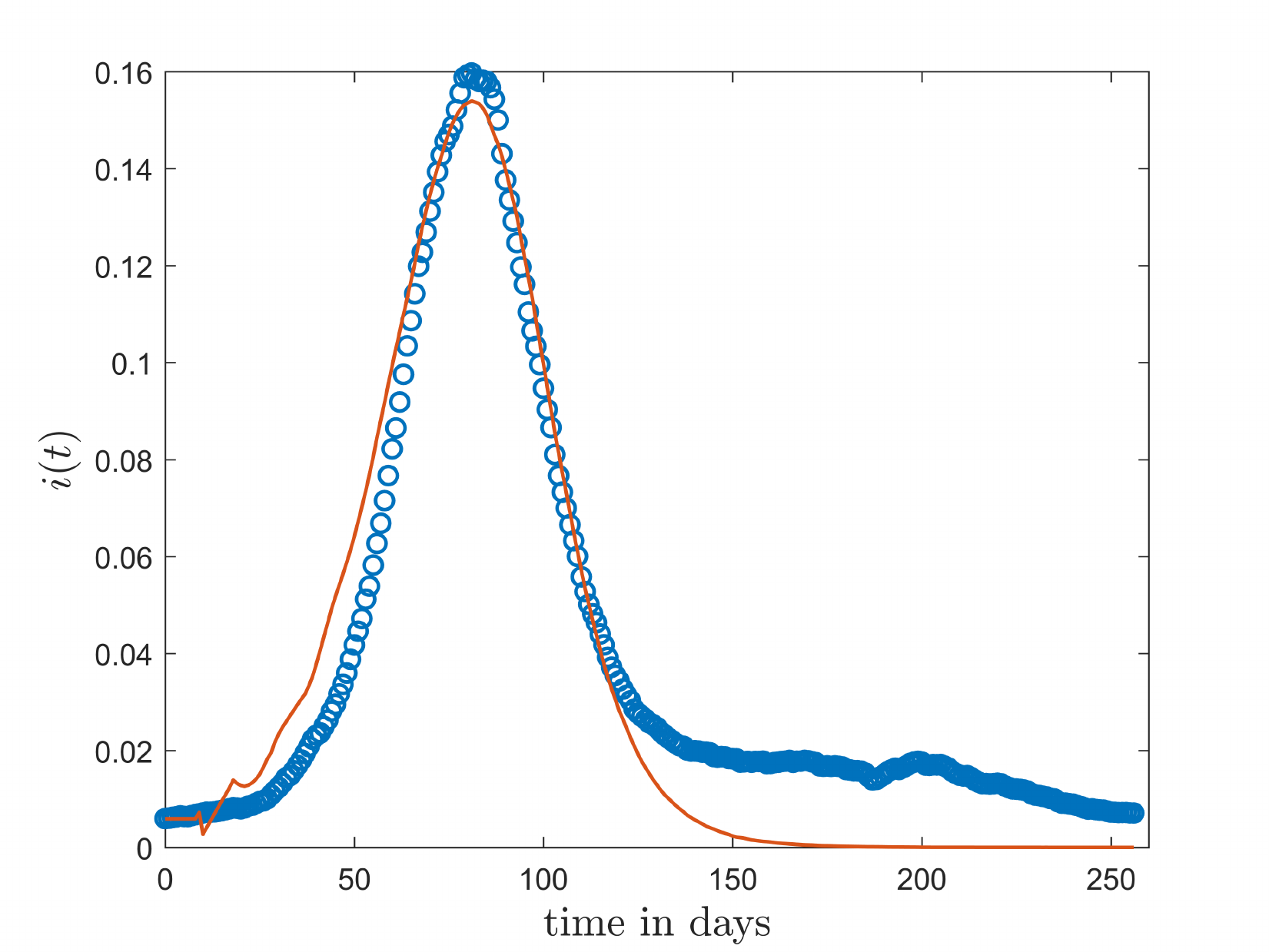}
	 		\caption{}
	 		\label{S_wave}
	 	\end{subfigure}
	 	\caption{{Model fitting of the COVID-19 data of India (a) Model fitting of the first wave. (b) Model fitting of the second wave. }\label{mod_fit}}	
	 \end{figure}

 	We have optimized COVID-19 data with a $101\times 101$ lattice space. Fig~\ref{F_wave} and \ref{S_wave} shows the fitted model along with data for the first and the second wave.
	 
	 Here we can see that the model fits reasonably well with the data. Also from the table~\ref{fitval_tab}, we see that the disease transmission probability ($q$) increases and the degree exponent ($n$) decreases in the second wave as compared to the first wave. Increment of $q$ represents that in the second wave of the disease it was possibly more infectious and spread faster than in the first wave.  Also decrement of $n$ indicates that the interaction between infectious and the susceptible population spread out to a larger distances in the second wave as compared to the first wave. This is possibly because of the COVID protocols which is relaxed much more in the initial phase of the second wave as compared to the first wave.
	 
\section{Conclusions}\label{conclusion}
	In this section we have summarized the main features and results of our model. The cellular automata (CA) is a very common tool to model a disease spread and has been used extensively in literature for studying different systems. In this paper, we have modeled the CA by proposing a new neighborhood criteria. Usually in earlier studies, neighborhood condition is such that the neighborhood of a lattice cell is always fixed. Whereas in our model, rather than choosing a specific neighborhood condition, we assume that a lattice cell can interact with any other cell at distance $d$ with a certain probability which is called interaction probability ($p_{int}(d)$) (Eq.\ref{pint_main}). We have assumed that the interaction probability ($p_{int}(d)$) is a function of the distance ($d$) and has a form of inverse power law with degree exponent $n$. Here, exponent $n$ is a very important parameter as it enables us to tune the social confinement of our model. With this newly defined neighborhood criteria, we have calculated various relations like average interaction distance ($\langle d\rangle$) and the probability of infection ($Q_{I}$) to understand and represent our model properly.
	
	From Fig.~\ref{avg_int_n}, we can see that the average interaction distance ($\langle d\rangle$) deceases and saturation is reached to $\sim 1$ as $n$ increases. So, for a higher $n$, a person can mainly interact with nearest neighbors. However, for a smaller value of $n$, a person can interact with the distant neighbors. Hence higher values of $n$ represents higher social confinement and vice-versa. Also we want to mention that for exponents, $n>3$, the average interaction distance, $\langle d\rangle\approx 1$. Thus the values $n\gg 3$ do not give us any new results.
	
	In the simulation section, we have studied the temporal and spatial behavior of our model for different degree exponents ($n$). As $n$ increases, the disease spread becomes slower and it is more clustered. This happens because of the decrease in the average interaction distance ($\langle d\rangle$) or in the other words increase in the social confinement with increasing values of $n$. Thus the disease transmission probability ($q$) and the degree exponent $n$ regulates the speed of the disease spread.
	
	Also, we have compared our model with COVID-19 data of India for different waves. We have first normalized the active cases of a wave with the total number of infected cases in that wave. Then we have optimized the sum of squared errors of the infectious cases (Eq.~\ref{SSE}) to get the best fit result with the data. Here all simulations are done on a $101\times 101$ lattice. We have found that the disease transmission probability ($q$) increases in the second wave than the first wave. This means that the disease is more infectious in the second wave than the first wave. Also the degree exponent ($n$) decreases in the second wave. This implies that the decrement in the COVID-19 restrictions (or decrement in the degree of social confinement) at the initial time of the second wave played a significant role to the faster spread of the disease. Our model fits the peak of the waves well, however fall in the data at the end of both waves and plateauing at the end of the second wave has some matches with our fitted model. This possibly indicates that our model needs to be modified to incorporate these aspects which we will look at in our future works.
	
	Modeling spread of a disease is a very complicated process since several factors have to be considered. Non-uniform distribution of the population and economic situation of the regions are two major factors which affects the disease spread. In future we want to look at these complex aspects by refining this model to find the behavior of the disease spreading more accurately. We would also like to study these possibilities in the context of COVID-19.
	
	\section*{Acknowledgment}
	
	 The authors would like to thank the Department of Physics, St. Xavier's College, Kolkata for providing support during this work. One of the authors (S. C.) acknowledges the financial support provided from the University Grant Commission (UGC) of the Government of India, in the form of CSIR-UGC NET-JRF. Finally, the authors would also like to express their gratitude to the anonymous referee for his/her valuable comments and suggestions.
	
	\bibliographystyle{unsrt}
	\bibliography{References}	
\end{document}